\documentclass[12pt, oneside, a4paper]{article}
\usepackage{mcite}
\usepackage{cite}
\usepackage{latexsym}
\usepackage{amssymb}
\usepackage{amsmath}
\usepackage{amsthm}
\usepackage[applemac]{inputenc}
\newcommand{\be}{\begin{eqnarray}}
            \newcommand{\ee}{\end{eqnarray}}
           \newcommand{\eel}[1]{\label{#1}\end{eqnarray}}
\newcommand{\e}[1]{\label{eq:#1}\end{eqnarray}}
     \newcommand{\eg}{{\em e.g.\ }}
            \newcommand{\ie}{{\em i.e.\ }}
            \newcommand{\ga}{{\gamma}}
 \newcommand{\Ga}{{\Gamma}}
            \newcommand{\la}{{\lambda}}
           \newcommand{\La}{{\Lambda}}  

\newcommand{\del}{{\delta}}

 \newcommand{\om}{{\omega}}

\newcommand{\dx}{{\dot{x}}}

\newcommand{\vt}{{\sf\scriptscriptstyle a}}
\newcommand{\sa}{{\sf\scriptscriptstyle sa}}
\newcommand{\st}{{\sf\scriptscriptstyle s}}
\newcommand{\gt}{{\sf\scriptscriptstyle g}}
\newcommand{\sgt}{{\sf\scriptscriptstyle sg}}
\newcommand{\DS}{{\sf\scriptscriptstyle ds}}
\newcommand{\sDS}{{\sf\scriptscriptstyle sds}}
\newcommand{\Lvt}{{\sf\scriptstyle a}}
\newcommand{\Lsa}{{\sf\scriptstyle sa}}
\newcommand{\Lst}{{\sf\scriptstyle s}}
\newcommand{\Lgt}{{\sf\scriptstyle g}}
\newcommand{\Lsgt}{{\sf\scriptstyle sg}}
\newcommand{\LDS}{{\sf\scriptstyle ds}}
\newcommand{\LsDS}{{\sf\scriptstyle sds}}

            \newcommand{\lra}{{\leftrightarrow}}
            
            \newcommand{\pet}{{\cal P}}

\newcommand{\de}{{\dot e}}
\newcommand{\dox}{{\dot x}}
\newcommand{\ddx}{{\ddot x}}
\newcommand{\dxi}{{\dot \xi}}

            \newcommand{\beq}{\begin{quote}}
            \newcommand{\eq}{\end{quote}}
            
            \newcommand{\al}{\alpha}
            \newcommand{\ben}{\begin{enumerate}}
            \newcommand{\een}{\end{enumerate}}
            \newcommand{\bit}{\begin{itemize}}
            \newcommand{\ei}{\end{itemize}}
        \newcommand{\nn}{\nonumber}
            \newcommand{\rl}[1]{(\ref{eq:#1})}
            \newcommand{\edfl}[1]{\Label{#1}\end{df}}

\newcommand{\vb}{{\cal h}}
\newcommand{\hb}{{\cal i}}

\newcommand{\tet}{{\theta}}
\newcommand{\dagg}{^{\dag}}

\newcommand{\dif}{{\partial}}
\newcommand{\half}{\frac{1}{2}}

\newcommand{\var}{\varepsilon}
\begin{document}
\begin{titlepage}
\vspace*{5 mm}
\vspace*{20mm}
\begin{center}
{\LARGE\bf Infinite spin particles}\end{center}
\vspace*{3 mm}
\begin{center}
\vspace*{3 mm}

\begin{center}Ludde Edgren\footnote{E-mail:
edgren@fy.chalmers.se}, Robert Marnelius\footnote{E-mail: 
tferm@fy.chalmers.se}
 and Per Salomonson\footnote{E-mail: tfeps@fy.chalmers.se}
 \\ \vspace*{7 mm} {\sl
Department of Fundamental Physics\\ Chalmers University of Technology\\
G\"{o}teborg University\\
S-412 96  G\"{o}teborg, Sweden}\end{center}
\vspace*{25 mm}
\begin{abstract}
We show that Wigner's infinite spin particle classically is described by a reparametrization invariant
higher order geometrical Lagrangian. The model exhibit unconventional
features like tachyonic behaviour and momenta proportional to light-like accelerations.
 A simple higher order superversion for half-odd integer particles is also derived. Interaction with external vector fields and curved spacetimes are  analyzed with negative results except for (anti)de Sitter spacetimes.  We 
quantize the free theories covariantly and show that the resulting wave functions are fields containing arbitrary 
large spins.  Closely related infinite spin particle models are also analyzed.\end{abstract}\end{center}\end{titlepage}

\setcounter{page}{1}
\setcounter{equation}{0}
\section{Introduction}
When Wigner
 \cite{Wigner:1939on,Bargmann:1948gr,Wigner:1963in} classified representations of the Poincar\'e group, he investigated the two Poincar\'e invariants $p_{\mu}p^{\mu}$ and $w_{\mu}w^{\mu}$ where $w^{\mu}$ is the Pauli-Lubanski vector defined by
 \be
 &&w^{\mu}\equiv\frac{1}{2}\var^{{\mu\nu\rho\sigma}}m_{{\nu\rho}}p_{\sigma},
 \e{01}
 where $m_{\mu\nu}$ and $p_{\mu}$ are the Poincar\'e generators and $\var^{{\mu\nu\rho\sigma}}$ the totally antisymmetric tensor. If $p_{\mu}$ is the four momentum of the particle, $p_{\mu}p^{\mu}$ is minus the mass squared ($p^2=-m^2$) for our choice of Minkowski metric (see below). For irreducible representations we have then $w^2=m^2s(s+1)$, where $s$ is the spin of the particle. For massless particles Wigner showed that apart from the natural representations, $p^2=w^2=0$, there are representations for which $p^2=0$ but $w^2=\Xi^2$, where $\Xi$ is a real, positive constant. These representations were called the continuous spin representation in \cite{Bargmann:1948gr} and the infinite spin representation in \cite{Wigner:1963in}. Wigner showed that it contains all helicities from $-\infty$ to $\infty$. In  \cite{Bargmann:1948gr} two representations were given in terms of covariant field equations: one for integer spins denoted $0(\Xi)$ (see also  \cite{Wigner:1963in}), and one for half-odd integer spins denoted $0'(\Xi)$. To our knowledge these representations have never been fully analyzed covariantly. In this paper we give therefore an extensive treatment starting from the original derivations, and in addition we construct classical geometrical particle models from the representations in terms of which we  investigate interactions with external fields including gravity. We also investigate quantum properties. We give a simple Gupta-Bleuler quantization of the free models which we believe to be in accordance with a correct BRST treatment.
 
 In section 2 we present some details for the representations given in
 \cite{Bargmann:1948gr,Wigner:1963in} and generalizations. In section
 3 we give  the classical theory corresponding to the main
 representations (the $\Xi$-representations). We show that they are naturally written in terms of
 a simple geometrical higher order Lagrangian. In section 4 we
 consider  super versions of the $\Xi$-representations and show that even
 here we have a natural higher order Lagrangian   involving
 odd Grassmann variables. In section 5 we then consider interactions
 with an external vector field and curved spacetimes. Although the
 results are negative we find consistent  models in (anti)de Sitter
 spacetimes provided $\Xi=0$. In view of the latter results we 
  give in section 6 some details of the modified models which only are possible when
 $\Xi=0$. In section 7 we quantize the various free
 models covariantly using a Gupta-Bleuler technique demonstrating  connections to higher spin fields. Finally
 in section 8 we conclude the paper. 
A complete survey of the constraints and their Poisson algebras
 considered in the text is given in appendices A and B.

\setcounter{equation}{0}
\section{Wigner's $\Xi$-representation}
In accordance with the treatment in references \cite{Bargmann:1948gr,Wigner:1963in} we consider a relativistic 
particle described by the coordinates $x^{\mu}$ and momenta 
$p_{\mu}$, $\mu=0,1,2,3$, together with an internal vector $\xi^\mu$ 
and its conjugate momentum $\pi_{\mu}$. The Lorentz' indices are 
raised and lowered by the spacelike Minkowski metric 
$\eta^{\mu\nu}$, $diag\, \eta^{\mu\nu}=(-1,+1,+1,+1)$. After quantization they 
satisfy the commutation relations (the nontrivial part)
\be
&&[x^\mu, p_{\nu}]=i\del^{\mu}_{\nu}, \qquad [\xi^\mu, \pi_{\nu}]=i\del^{\mu}_{\nu}.
\e{1}
The Poincar\'{e} generators are  $m_{\mu\nu}$ and $p^{\mu}$, where $m_{\mu\nu}$ are
the generators of the Lorentz' transformations  given by
\be
&&m^{\mu\nu}=l^{\mu\nu}+s^{\mu\nu},\nn\\
&&l^{\mu\nu}=x^\mu p^\nu -x^\nu p^\mu, \qquad
s^{\mu\nu}=\xi^\mu\pi^\nu-\xi^\nu\pi^\mu.
\e{2}
Following Wigner \cite{Wigner:1939on,Bargmann:1948gr,Wigner:1963in} we classify the 
representations by the properties of the Poincar\'{e} invariants 
$p^2$ and $w^2$ where
$p_{{\mu}}$ is the momentum of the particle and $w^\mu$ the Pauli-Lubanski vector \rl{01}
for which we have the relation
\be
&&w^{\mu}\equiv\frac{1}{2}\var^{{\mu\nu\rho\sigma}}m_{{\nu\rho}}p_{\sigma}=
\frac{1}{2}\var^{{\mu\nu\rho\sigma}}s_{{\nu\rho}}p_{\sigma}.
\e{3}
($\var^{{\mu\nu\rho\sigma}}$ is the totally antisymmetric tensor 
with $\var^{0123}=1$.) Notice that  {$p_{\mu}w^{\mu}\equiv0$}. In particular 
we are interested in the massless representations where $p^2$ is zero. This we do by looking for a physical subspace of the complete state 
space in which $p^2$ is zero and $w^2$ a 
constant. In strong form these conditions may be written as
\be
&&p^2|phys\hb=0, \qquad w^2|phys\hb=\Xi^2|phys\hb,
\e{4}
where
$\Xi$ is the real, positive constant introduced by Wigner \cite{Wigner:1939on,Bargmann:1948gr,Wigner:1963in}.
Using the equality
\be
&&w^2=-\half
s_{\mu\nu}s^{\mu\nu}p_{\rho}p^\rho-s_{\mu\nu}s^{\nu\la}p_{\la}p^\mu,
\e{5}
and the property ($p\cdot\pi\equiv p_{\mu}\pi^{\mu}$ etc.)
\be
&-s_{\mu\nu}s^{\nu\la}p_{\la}p^\mu=&2p^2+\pi^2(p\cdot\xi)^2+\xi^2(p\cdot\pi)^2-(\pi\cdot\xi)(p\cdot\xi)(p\cdot\pi)-\nn\\&&-(\pi\cdot\xi)(p\cdot\pi)(p\cdot\xi),\nn\\
\e{6}
 we find two natural sets of elementary constraints in the chosen variables that solves \rl{4}. The first condition in \rl{4} is already elementary and \rl{6} yields then  from the second condition in \rl{4} either i) $(p\cdot\xi)|phys\hb=0$, or ii) $(p\cdot\pi)|phys\hb=0$. In the first case the second equation in \rl{4} reduces to
\be
&&\xi^2(p\cdot\pi)^2|phys\hb=\Xi^2|phys\hb,
\e{7}
from which we finally get the sufficient minimal set of constraints to be
\be
&&\chi_1|phys\hb=\chi_2|phys\hb=\chi_3|phys\hb=\chi_4|phys\hb=0,
\e{8}
where
\be
&&\chi_1\equiv\half p^2,\qquad\chi_2\equiv\half\bigl(\xi^2-F^2\bigr),\nn\\
&&\chi_3\equiv p\cdot\xi,\qquad\chi_4\equiv p\cdot\pi-{\Xi\over F}.
\e{9}
The factors one-half are chosen for convenience (see next section). $F$ is a nonzero constant (or a nonzero operator commuting with $p, x, \pi, \xi$, see later). It may be fixed by a rescaling of  $\pi, \xi$.   In the second case we find similarly the sufficient set of constraints to be \rl{8} where
\be
&&\chi_1\equiv\half p^2,\qquad\chi_2\equiv\half\bigl(\pi^2-F^2\bigr),\nn\\
&&\chi_3\equiv p\cdot\pi,\qquad\chi_4\equiv p\cdot\xi-{\Xi\over F}.
\e{10}
For $F=1$ the representation \rl{9} is exactly the one given in \cite{Bargmann:1948gr,Wigner:1963in}. The representation \rl{10} is essentially equivalent to \rl{9} since only $\xi$ and $\pi$ are interchanged.

In the derivation of such explicit representations as the above ones there are two properties we must secure: that the complete set of constraints is hermitian, and that they satisfy a closed commutator algebra, \ie
\be
&&[\chi_k, \chi_l]=iC_{klm}\chi_m,
\e{11}
which makes \rl{8} consistent. The choices \rl{9} and \rl{10} satisfy these criteria and yield $C_{klm}$ that are constants. The resulting theory is therefore a gauge theory for which \rl{11} is the Lie algebra of the gauge group. Explicitly we have here the Lie algebra (the nonzero part)
\be
&&[\chi_4, \chi_2]=i\chi_3,\qquad [\chi_4, \chi_3]=2i\chi_1,
\e{1101}
for \rl{10}, and the same algebra with minus signs for \rl{9}.
It is possible to contemplate weaker conditions then \rl{8} if the representations are derived from the weak condition $\vb phys|(w^2-\Xi^2)|phys\hb=0$.
Such weaker conditions are natural within a BRST framework and will be considered in section 7.

Going back to our derivations above one may notice that in the case when $\Xi=0$ it is sufficient to impose
\be
&&p^2|phys\hb=0,\qquad(p\cdot\xi)|phys\hb=0,\qquad (p\cdot\pi)|phys\hb=0,
\e{12}
in order to satisfy $w^2|phys\hb=0$. On the other hand we may always add further constraints like $(\xi^2-C^2)|phys\hb=0$, or  $(\pi^2-C^2)|phys\hb=0$, where $C$ is any  constant (\eg one) without violating the criteria above, \ie the constraints still satisfy \rl{11}. This case will be further treated in section 6.

\setcounter{equation}{0}
\section{Classical model for the $\Xi$-representation}
In this section we treat $p, x$ and $\pi, \xi$ as classical variables satisfying the Poisson bracket relations (the non-zero part)
\be
&&\{x^{\mu}, p_{\nu}\}=\del^{\mu}_{\nu},\qquad \{\xi^{\mu}, \pi_{\nu}\}=\del^{\mu}_{\nu}.
\e{301}
They are furthermore viewed as dynamical functions of a time parameter $\tau$. In terms of these variables we define the  Hamiltonian of this infinite spin particle model   to be
\be
&&H\equiv \la_1\chi_1+ \la_2\chi_2+ \la_3\chi_3+ \la_4\chi_4,
\e{302}
where  $\la_i$ are Lagrange multipliers and where $\chi_i$ are the constraint variables defined to be the classical counterparts of the operators in \rl{9} or \rl{10}. 
The Lagrangian is then obtained in its phase space form through the Legendre transformation, \ie
\be
&&L=p_{\mu}\dox^{\mu}+\pi_{\mu}\dxi^{\mu}-H,
\e{303}
where $\dox\equiv{dx\over d\tau}$ etc (in \rl{301}-\rl{303} all variables are given for the same value of $\tau$).
Under certain conditions on the Lagrange multipliers we may express $L$ in terms of $x$ and $\xi$ only.
In the case in which $\chi_i$ are given by \rl{9} with $F=1$ we thus find the Lagrangian in configuration space to be
\be
&&L=-{\la_1\over 2\la_4^2}\dxi^2+{1\over\la_4}(\dox-\la_3\xi)\cdot\dxi-\half\la_2(\xi^2-1)+\la_4\Xi,\quad\la_4\neq0.
\e{304}
In the case in which $\chi_i$ are given by \rl{10} with the choice $F=1$ the configuration Lagrangian is given by
\be
&&L={\la_2\over 2A}(\dox-\la_4\xi)^2+{\la_1\over2A}\dxi^2-{\la_3\over A}(\dox-\la_4\xi)\cdot\dxi+\half\la_2+\la_4\Xi,\nn\\&& A\equiv\la_1\la_2-\la_3^2\neq0.
\e{305}
 Notice that these theories are gauge theories since the constraint variables satisfy a Lie algebra in terms of the Poisson bracket \rl{301}. We have  (the nonzero part)
 \be
&&\{\chi_4, \chi_2\}=\chi_3,\qquad \{\chi_4, \chi_3\}=2\chi_1,
\e{3051}
for \rl{10}, and the same algebra with minus signs for \rl{9}. These algebras are nilpotent.

One may notice that the constraints in configuration space following from \rl{304} and \rl{305} are different for whatever choice we make of the Lagrange multipliers. This implies that the variable $\xi^{\mu}$ has different meanings in the two cases. In order to find a simple more geometrical Lagrangian we try to eliminate $\xi^{\mu}$ by means of its  equation of motion,
\be
&&{d\over d\tau}{\dif L\over\dif\dxi^{\mu}}-{\dif L\over\dif\xi^{\mu}}=0.
\e{306}
We notice then that this equation for the Lagrangian \rl{304} reduces to
\be
&&\xi^{\mu}=-{1\over\la_2}{d\over d\tau}\biggl({1\over\la_4}\dox^{\mu}\biggr),
\e{307}
provided $\la_1=0$, $\la_3=\al\la_4$, and $\la_2\neq0$ for arbitrary real constants $\al$. When \rl{307} is inserted back into \rl{304} we find apart from total derivatives
\be
&&L={1\over2\la_2}\biggl({d\over d\tau}\bigl({1\over\la_4}\dox^{\mu}\bigr)\biggr)^2+\half\la_2+\la_4\Xi.
\e{308}
Although this Lagrangian  have only two Lagrange multipliers, a Dirac consistency check using the equations of motion will generate the complete set of  constraints (see below). 

If we instead make use of the Lagrangian \rl{305} in \rl{306}, the latter reduces to
\be
&&\xi^{\mu}={1\over\la_4}\dox^{\mu}-{\al\over\la_2}{d\over d\tau}\biggl({1\over\la_4}\dox^{\mu}\biggr),
\e{309}
provided  $\la_1=0$, $\la_3=\al\la_4$, and $\la_2\neq0$ for any real $\al\neq0$. When \rl{309} is inserted back into \rl{305} we also here find  the Lagrangian \rl{308}. 

We have arrived at the unique Lagrangian \rl{308} as a classical model for Wigner's $\Xi$-representation. One may notice that even the Lagrange multiplier $\la_2$ may be eliminated from \rl{308} in which case \rl{308} reduces to
\be
&&L=\sqrt{\biggl({d\over d\tau}\bigl({1\over\la_4}\dox\bigr)\biggr)^2}+\la_4\Xi.
\e{310}
This is the  most simple and geometrical form of the Lagrangian since $\la_4$ cannot be eliminated.
Notice that it represents a reparametrization invariant theory where $\la_4$ is the einbein variable (usually denoted $v$).

\subsection{Analysis of the geometrical Lagrangian}
The geometrical Lagrangian \rl{310} is a higher order Lagrangian since it involves the second derivative of $x^{\mu}$. To analyze its properties  is therefore nontrivial. This analysis is made slightly more convenient if we  write the  Lagrangian \rl{310} in terms of  
 the inverse einbein variable, $e=1/\la_4$. We have then
\be
&&L=\sqrt{\biggl({d\over d\tau}\bigl(e\dox\bigr)\biggr)^2}+{1\over e}\Xi\equiv\sqrt{\bigl(\de\dox+e\ddx\bigr)^2}+{1\over e}\Xi.
\e{311}
In order to transform this theory into the Hamiltonian framework we must make use of Ostrodgradski's method \cite{Ostrogradski:1850mv} (see also \cite{Whittaker:1937an} chapter X, or better \cite{Lanczos:1970va} appendix I). This method requires us to introduce a new variable. Even though it might cause confusion we call also this variable $\xi^{\mu}$ since it  is similar although not identical to the variable used before. Here  it is defined by
\be
&&\xi^{\mu}\equiv\dox^{\mu}.
\e{312}
Ostrodgradski requires us then to replace  $\dx^{\mu}$ and $\ddx^{\mu}$ in $L$ by $\xi^{\mu}$ and $\dxi^{\mu}$ and to define the Hamiltonian  by
\be
&&H=p_{\mu}\xi^{\mu}+\pi_{\mu}\dxi^{\mu}+\om\de-L(\xi,\dxi,e,\de),
\e{313}
where $p_{\mu}$ as before is the conjugate momentum to $x^{\mu}$.  $\pi_{\mu}$ is the conjugate momentum to $\xi^{\mu}$, and $\om$ is the conjugate momentum to $e$ which is necessary here since $L$ contains $\dot{e}$. The derivatives $\dxi^{\mu}$ and $\de$ are eliminated from $H$ by the equalities
\be
&&\pi_{\mu}={\dif L\over\dif\dxi^{\mu}}={e(e\dxi_{\mu}+\de\xi_{\mu})\over\sqrt{(e\dxi+\de\xi)^2}},
\e{314}
\be
&&\om={\dif L\over\dif\de}={(e\dxi\cdot\xi+\de\xi^2)\over\sqrt{(e\dxi+\de\xi)^2}}.
\e{315}
These equalities inserted into \rl{313} yields then
\be
&&H=p\cdot\xi-{1\over e}\Xi.
\e{316}
However, in addition they yield the following primary constraints (the constraints are numbered in accordance with \rl{10})
\be
&&2\chi_2\equiv\pi^2-e^2,
\e{317}
\be
&&\chi_5\equiv\pi\cdot\xi-\om e.
\e{318}
The total Hamiltonian, $H_{tot}$, which governs the time evolution, is obtained by adding  a linear combination of these primary constraints to  \rl{316}. We have
\be
&&H_{tot}=H+\la_2\chi_2+\la_5\chi_5.
\e{3181}
The Poisson bracket is here (the nontrivial part)
\be
&&\{x^{\mu}, p_{\nu}\}=\del^{\mu}_{\nu},\qquad\{\xi^{\mu}, \pi_{\nu}\}=\del^{\mu}_{\nu},\qquad\{e, \om\}=1,
\e{319}
from which we then get the equations
\be
&&\dot{\chi}_2=\{\chi_2, H_{tot}\}=-p\cdot\pi+2\la_5\chi_2,\nn\\
&&\dot{\chi}_5=\{\chi_5, H_{tot}\}=-p\cdot\xi+{1\over e}\Xi-2\la_2\chi_2\equiv-H-2\la_2\chi_2.
\e{320}
 Since we have to impose the conditions $\dot{\chi}_2=0$, $\dot{\chi}_5=0$ for consistency we arrive at the secondary constraints $\chi_3=0$ and $\chi_4=0$ where
\be
&&\chi_3\equiv p\cdot\pi,\qquad\chi_4\equiv H=p\cdot\xi-{1\over e}\Xi,
\e{321}
which is consistent with the fact that the Hamiltonian always is zero in a reparametrization invariant theory.  Furthermore, we find
\be
&&\dot{\chi}_3=\{\chi_3, H_{tot}\}=-p^2-\la_5\chi_3,\nn\\&&\dot{\chi}_4=\{\chi_4, H_{tot}\}=\la_2\chi_3+\la_5\chi_4,
\e{322}
from which we by consistency have to impose the tertiary constraint $\chi_1=0$ where
\be
&&2\chi_1\equiv p^2.
\e{323}
There are no further constraints since $\{p^2, H_{tot}\}=0$. Thus, the Lagrangian \rl{311} gives rise to five constraints although it was derived from a Hamiltonian, \rl{302}, involving only  four. The reason is that the einbein variable has become dynamical after we eliminated $\xi^{\mu}$. A new constraint is therefore necessary in order to remove the new degree of freedom. We notice that all five constraints satisfy a Lie algebra. In fact, they satisfy the algebra \rl{3051} together with
\be
&&\{\chi_5, \chi_2\}=2\chi_2,\quad \{\chi_5, \chi_4\}=-\chi_4,\quad\{\chi_5, \chi_3\}=\chi_3,
\e{324}
which is the Lie algebra of the gauge group (see also appendix A.1). This algebra is solvable. It is the constraint $\chi_5$ in \rl{318} which is new here. By means of the gauge choice $e=1$, $\chi_5$ is eliminated ($\{\chi_5, e-1\}=e=1\neq0$) and we are left with exactly the constraints \rl{10} with $F=1$ used before. Notice also that the five constraints here are exactly given by \rl{10} with $F=e$ together with $\chi_5$ in \rl{318}. The careful reader may also note another puzzling feature: when we eliminated $\xi^{\mu}$ by means of \rl{307} or \rl{309} we actually removed two constraints by our choice of Lagrange multipliers. The reason why these constraints are recovered is that Dirac's consistency conditions bring them back.

The $\xi^{\mu}$ variable used in this section is different from the one used before. Comparing the expressions \rl{312} and \rl{309} with the $\xi^{\mu}$ used here we notice that they differ by an acceleration term and a rescaling. This rescaling is reflected in the form of the constraints here. There is, however, no resemblance between \rl{312} and \rl{307} which is consistent with the fact that the constraint $\chi_2$ in \rl{317} is not contained in the set \rl{9}.

Concerning the meaning of this particle model we notice that the constraint $\chi_4=0$ from \rl{321} implies that $\dot{x}^{\mu}$ is spacelike for $e>0$ and $\Xi\geq0$ with $p^0>0$ and $\dot{x}^0>0$ ($p^{\mu}\propto\dot{x}^{\mu}$ is excluded by the other constraints). For $\Xi<0$ we have no definite sign of ${\dot{x}}^2$. Thus, the model can possibly exhibit non-tachyonic features by choosing $\Xi<0$, but in general we have tachyonic behavior of the particle.

\subsection{Remarks}
Choosing $e=1$ in the Lagrangian \rl{311} we find
\be
&&L=\sqrt{\ddx^2}+\Xi,
\e{325}
which apart from the constant $\Xi$ is the Lagrangian considered by Zoller \cite{Zoller:1994cl}. Notice, however, that the constraints following from \rl{325} are not the same, even for $e=1$. We get using Ostrogradski's method the following three constraints
\be
&&\pi^2-1=0,\qquad p\cdot\pi=0, \qquad p^2=0.
\e{326}
Although a time independent quantization yields
\be&&
p\cdot\xi={\rm const},
\e{327}
 this constant is not fixed but is related to the energy spectrum $E$. For the model \rl{325} it is $\Xi+E$. In \cite{Zoller:1990in} Zoller proposes a reparametrization invariant version of \rl{325} which also yields five constraints. However, his model is entirely different from our model. His constraints are both inconsistent with ours and with those in \rl{326} and \rl{327}.

The generalizations of the actions \rl{311}  and \rl{325} to string theory have been considered by Savvidy \cite{Savvidy:2003co} (see also \cite{Savvidy:2003te,*Antoniadis:2004ph,*Nichols:2002ne}). The generalization of \rl{310} to arbitrary dimensions is
\be
&&S=\int d^m\zeta\biggl(\sqrt{h}\sqrt{(\triangle(h)X^{\mu})^2}+\Xi\sqrt{h}\biggr),\quad h=\det h_{ab},
\e{328}
where $h_{ab}$ is the metric on the manifold coordinatized by $\zeta^a$. $\triangle(h)$ is the Laplace-Beltrami operator $\bigl(1/\sqrt{h}\bigr)\dif_a\sqrt{h}h^{ab}\dif_b$ where $\dif_a\equiv{\dif\over\dif\zeta^a}$. For $\Xi=0$ and $m=2$ \rl{328} is exactly the model B in \cite{Savvidy:2003co}.

\setcounter{equation}{0}
\section{Superextended model}
Let us add to the previous operators in section 2 the odd operator $\psi^{\mu}$ satisfying the commutation relations (the nonzero ones)
\be
&&[\psi^{\mu}, \psi^{\nu}]_+=\eta^{\mu\nu},
\e{401}
where the index plus indicates an anticommutator.  $\psi^{\mu}$ is an odd, hermitian operator that transforms as a Lorentz vector. The Lorentz generators are here (cf.\rl{2})
\be
&&m^{\mu\nu}=l^{\mu\nu}+s^{\mu\nu},\qquad
l^{\mu\nu}=x^\mu p^\nu -x^\nu p^\mu, \nn\\&&
s^{\mu\nu}=\xi^\mu\pi^\nu-\xi^\nu\pi^\mu-{i\over2}\bigl(\psi^{\mu}\psi^{\nu}-\psi^{\nu}\psi^{\mu}\bigr).
\e{402}
The conditions \rl{4} for Wigner's $\Xi$-representation yield then
\be
&&p\cdot\psi|phys\hb=0
\e{403}
together with  the previous constraints \rl{9} or \rl{10}   as a sufficient minimal elementary set of restrictions. Note that
\be
&&[p\cdot\psi, p\cdot\psi]_+=p^2
\e{404}
In a wave function representation \rl{403} is a Dirac equation (the gamma matrices may then be identified with $\sqrt{2}\psi^{\mu}$). In this form this set of constraints were also given in \cite{Bargmann:1948gr} as a representation of the half-odd integer case denoted $0'(\Xi)$.

\subsection{Pseudoclassical model}
If we only add the constraint $\chi_6|phys\hb=0$ where
\be
&&\chi_6\equiv p\cdot\psi
\e{405}
to the previous constraints in \rl{9} or \rl{10} we are unable to obtain a simple Lagrangian at the classical level, particularly not a higher order one. However, if we also introduce the odd, hermitian operator $\tet$ satisfying the anticommutation relation
\be
&&[\tet, \tet]_+=-1,
\e{406}
together with the new constraint (in conjunction with \rl{10} for $F=1$)
\be
&&(\pi\cdot\psi+\tet)|phys\hb=0
\e{407}
it is possible to construct a simple, higher order pseudoclassical model which roughly contains the original model (see below).

At the pseudoclassical level we have then  the real, odd variables $\psi^{\mu}$ and $\tet$ satisfying the (super) Poisson bracket relations (the nonzero part)
 \be
 &&\{\psi^{\mu}, \psi^{\nu}\}=-i\eta^{\mu\nu}, \quad \{\tet, \tet\}=i,
 \e{408}
 together with \rl{301}.
 We  consider then apart from \rl{10} the constraint variables (we
 insert an index $\Lst$ on all constraints for this model even though the previous set \rl{10} are not changed) 
 \be
 &&\chi^\st_6\equiv p\cdot\psi,\quad\chi^\st_7\equiv\pi\cdot\psi+\tet.
 \e{409}
 Together with \rl{10} for $F=1$ they satisfy a Lie algebra whose nonzero part is given by \rl{3051} and
 \be
 &&\{\chi^\st_6, \chi^\st_6\}=-2i\chi^\st_1,\quad\{\chi^\st_7, \chi^\st_7\}=-2i\chi^\st_2,\nn\\
 &&\{\chi^\st_6, \chi^\st_7\}=-i\chi^\st_3,\quad\;\;\{\chi^\st_7, \chi^\st_4\}=-\chi^\st_6.
 \e{410}
 The constraint $\chi^\st_7$ seems to be possible to eliminate by a gauge condition on $\tet$ since $\{\chi^\st_7,\tet\}=i\neq0$, which then would leave the minimal set of constraints considered above and in \cite{Bargmann:1948gr}. However, this equivalence is not entirely correct since such an elimination of $\tet$ also would reduce the degrees of freedom of $\psi^{\mu}$. 
 
 As before we start our analysis from an extended Hamiltonian, here given by
\be
&&H^\st= \la_1\chi^\st_1+ \la_2\chi^\st_2+ \la_3\chi^\st_3+ \la_4\chi^\st_4+i\la_6\chi^\st_6+i\la_7\chi^\st_7,
\e{411}
 where $\la_6$ and $\la_7$ are new odd, real Lagrange multipliers.
 The Lagrangian in phase space is then
\be
&&L^\st=p_{\mu}\dox^{\mu}+\pi_{\mu}\dxi^{\mu}+{i\over 2}\psi\cdot\dot{\psi}-{i\over 2}\tet\dot{\tet}-H^\st.
\e{412}
In the case in which $\chi^\st_i$, $i=1,2,3,4,$ are given by \rl{10} with the choice $F=1$ the configuration Lagrangian is given by ($\psi^{\mu}$ and $\tet$ are really phase space variables)
\be
&L^\st= &{i\over 2}\psi\cdot\dot{\psi}-{i\over 2}\tet\dot{\tet}+\half\la_2+\la_4\Xi-i\la_7\tet+       {1\over A}\biggl(\half\la_2\dox^2+\half\la_1\dxi^2-\la_3\dox\cdot\dxi-\nn\\&&-\la_2\la_4\dox\cdot\xi+\la_3\la_4\xi\cdot\dxi+\half\la_2\la_4^2\xi^2-\la_2\la_6i\dox\cdot\psi+\la_3\la_7i\dox\cdot\psi+\nn\\&&+\la_3\la_6i\dxi\cdot\psi-\la_1\la_7i\dxi\cdot\psi+\la_2\la_4i\la_6\xi\cdot\psi-\la_3\la_4i\la_7\xi\cdot\psi\biggr),\nn\\&&A\equiv\la_1\la_2-\la_3^2\neq0.
\e{413}
This is a gauge theory since the constraint variables satisfy a Lie algebra. Under the same conditions as before on the Lagrange multipliers $\la_i$ we may eliminate $\xi^{\mu}$ from $L^\st$. We choose also to impose $\la_6=0$ which is possible due to the presence of $\chi_7$. The equation of motion \rl{306} yields then
\be
&&\xi^{\mu}={1\over\la_4}\dox+{\al\over\la_2}\biggl(i\la_7\psi^{\mu}-{d\over d\tau}\bigl({1\over\la_4}\dox\bigr)\biggr),
\e{414} 
for $\la_1=0$, $\la_3=\al\la_4$, $\la_2\neq0$ and $\la_6=0$. When \rl{414} is inserted into \rl{413} with the same conditions on the Lagrange multipliers we find
\be
&L^\st=&{1\over2\la_2}\biggl({d\over d\tau}\bigl({1\over\la_4}\dox\bigr)\biggr)^2+\half\la_2+\la_4\Xi-{i\la_7\over\la_2}\psi\cdot\biggl({d\over d\tau}\bigl({1\over\la_4}\dox\bigr)\biggr)+\nn\\&&+{i\over 2}\psi\cdot\dot{\psi}-{i\over 2}\tet\dot{\tet}-i\la_7\tet.
\e{415}
After eliminating $\la_2$ we  finally get
\be
L^\st&\!=&\!\sqrt{\biggl({d\over d\tau}\bigl(e\dox\bigr)\biggr)^2-2i\la_7\psi\cdot{d\over d\tau}\bigl(e\dox\bigr)}+{1\over e}\Xi+{i\over 2}\psi\cdot\dot{\psi}-{i\over 2}\tet\dot{\tet}-i\la_7\tet\equiv\nn\\&\!\equiv&\!\sqrt{\bigl(\de\dox+e\ddx\bigr)^2-2i\la_7\psi\cdot\bigl(\dot{e}\dox+e\ddx\bigr)}+{1\over e}\Xi+{i\over 2}\psi\cdot\dot{\psi}-{i\over 2}\tet\dot{\tet}-i\la_7\tet,\nn\\
\e{417}
where $e$ as before is the inverse einbein variable given by $e=1/\la_4$. (Notice that the expression under the square root sign may be written as $A^\mu A_\mu$ where
$A^\mu\equiv \de\dox^{\mu}+e\ddx^{\mu}-i\la_7\psi^{\mu}$.)
Applying Ostrogradski's method to the Lagrangian \rl{417} we obtain the Hamiltonian
\be
&&H^\st=p_{\mu}\xi^{\mu}+\pi_{\mu}\dxi^{\mu}+\om\de+{i\over 2}\psi\cdot\dot{\psi}-{i\over 2}\tet\dot{\tet}-L^\st(\xi,\dxi,e,\de,\psi,\dot{\psi},\la_7),\nn\\
\e{418}
where as before $\xi^{\mu}=\dox^{\mu}$ etc. We have here
\be
&&\pi_{\mu}={\dif L^\st\over\dif\dxi^{\mu}}={e(e\dxi_{\mu}+\de\xi_{\mu})-i\la_7e\psi_{\mu}\over\sqrt{(e\dxi+\de\xi)^2-2i\la_7\psi\cdot\bigl(e\dxi+\de\xi\bigr)}},
\e{419}
\be
&&\om={\dif L^\st\over\dif\de}={(e\dxi\cdot\xi+\de\xi^2)-i\la_7\psi\cdot\xi\over\sqrt{(e\dxi+\de\xi)^2-2i\la_7\psi\cdot\bigl(e\dxi+\de\xi\bigr)}}.
\e{420}
From these relations we find the equality
\be
&&\pi_\mu\dxi^{\mu}+\om\de-i\la_7{1\over e}\pi_\mu\psi^\mu=\sqrt{(e\dxi+\de\xi)^2-2i\la_7\psi\cdot\bigl(e\dxi+\de\xi\bigr)}
\e{421}
which when
inserted into \rl{418} yields  the Hamiltonian (we insert an index $(\la)$ to indicate its dependence on $\la_7$)
\be
&&H^\st_{(\la)}=p_{\mu}\xi^{\mu}-{1\over e}\Xi +i\la_7\bigl({1\over e}\pi\cdot\psi+\tet\bigr).
\e{422}
In addition we get the primary constraints $\pet_7=0$,~ $\chi^\st_2=0$ and $\chi^\st_5=0$, where $\pet_7$ is the conjugate momentum to $\la_7$ and
\be
&&2\chi^\st_2\equiv\pi^2-e^2,
\e{423}
\be
&&\chi^\st_5\equiv\pi\cdot\xi-\om e.
\e{424}
(We ignore the trivial constraints $\pet_i=0$ in general, where $\pet_i$ are conjugate momenta to $\la_i$. However, here $\pet_7=0$ is necessary to include due to the $\la_7$-dependence in $H^\st_{(\la)}$.)
The time evolution is  determined by the total Hamiltonian
\be
&&H^\st_{tot}=H^\st_{(\la)}+\la_2\chi^\st_2+\la_5\chi^\st_5+i\rho\pet_7,
\e{4251}
where $\rho$ is a real, odd Lagrange multiplier.
The consistency conditions $\dot{\chi^\st_2}=0$,  $\dot{\chi^\st_5}=0$, and  $\dot{\pet_7}=0$ yield then the secondary constraints (we are using the Poisson bracket defined by \rl{319} and \rl{408})
\be
&&\chi^\st_3\equiv p\cdot\pi,\quad\chi^\st_4\equiv p\cdot\xi-{1\over e}\Xi,\nn\\
&&\chi^\st_6\equiv p\cdot\psi,\quad\chi^\st_7\equiv\pi\cdot\psi+e\tet.
\e{426}
(As in section 3, $\chi^\st_4$ may be identified with the elementary Hamiltonian given in \rl{316} (cf.\rl{321}).)
Further derivations with respect to $\tau$ yield finally the tertiary constraint $\chi^\st_1=0$ given in \rl{323}. Since no further constraints are required we end up with seven constraints plus the trivial $\pet_7=0$. They satisfy the Lie algebra \rl{3051}, \rl{324}, \rl{410}, and
\be
&&\{\chi^\st_5, \chi^\st_6\}=0,\qquad \{\chi^\st_5, \chi^\st_7\}=\chi^\st_7.
\e{427}
($\pet_7$ yields zero Poisson brackets with $\chi^\st_i$.)
Again it is the dynamical einbein variable which has caused the appearance of
 the new constraint $\chi^\st_5$ although we started without it. Furthermore, we removed three constraints, $\chi_1^\st$, $\chi_3^\st$, $\chi_6^\st$, when we eliminated $\xi^{\mu}$ by \rl{414}, which however are recovered by Dirac's consistency conditions in the final model.
 
%
%
 \setcounter{equation}{0}
\section{Interactions with external fields}
In this section we consider interactions with some external classical 
fields at the classical level. The interactions are chosen to be
consistent with reparametrization invariance.  We consider both the
ordinary case given in section 3 and the pseudoclassical case given in
section 4.  We start with the ordinary case. (In each model below, the Lagrangian, the Hamiltonian and the constraints are identified with a specific index. Constraints and their Poisson algebras for each model are also listed in appendices A and B.)
 \subsection{Interaction with a vector field (index $\Lvt$)}
Like in the case of the ordinary relativistic particle there is a natural choice for a reparametrization invariant  interaction with an
external vector potential $A^{\mu}(x)$. It is given by
\be
&&L^\vt= \sqrt{(e\ddot{x}+\dot{e}\dot{x})^2} + {1\over e}\Xi 
+ A_{\mu}(x)\dot{x}^{\mu}.
\e{501}
 Within Ostogradski's procedure the interaction term $A_{\mu}(x)\dot{x}^{\mu}$
is replaced by $A_{\mu}(x)\xi^{\mu}$ which contains no $\tau$ derivative.
Therefore the expressions for the conjugate momenta and the
primary constraints are the same as for the free model except that $p_{\mu}$ is replaced by
 \be
 &&\La^\vt_{\mu}\equiv p_{\mu}-A_{\mu}(x).
 \e{5011} 
 Thus, we have the standard minimal coupling. The Hamiltonian is \eg
 turned into  
\be
&&H^\vt = \La^\vt\cdot\xi-{1\over e}\Xi.
\e{502}
The primary constraints $\chi^\vt_2$, $\chi^\vt_5$, are unchanged
(i.e.~$\chi^\vt_2\equiv\chi_2$, $\chi^\vt_5\equiv\chi_5$), but the 
expressions for the secondary ones are modified to
\be
&&\chi^\vt_3\equiv\pi\cdot\La^\vt,\qquad
 \chi^\vt_4 \equiv H^\vt,
\e{504}
 and the tertiary constraint $\chi^\vt_1$ becomes here
 \be
 && 2\chi^\vt_1\equiv(\La^\vt)^2+\half F_{\mu\nu}(x)(\xi^{\mu}\pi^{\nu}-\xi^{\nu}\pi^{\mu}),
 \e{5041}
 where $F_{\mu\nu}$ is the field strength defined by
 \be
 &&F_{\mu\nu}\equiv\dif_{\mu}A_{\nu}-\dif_{\nu}A_{\mu}.
 \e{5042}
Notice the following Poisson bracket relations for the new variable $\Lambda^\vt$
\be
&&\{x^\mu,\La^\vt_\nu\}=\del^\mu_{\;\nu},\qquad \{\La^\vt_\mu,\La^\vt_\nu\}=F_{\mu\nu}(x).
\e{503}
 However, apart from the generalized free constraints the consistency condition $\dot{\chi}^\vt_1=0$  requires the vanishing of the expression
\be
&&2\{\chi^\vt_1,\chi^\vt_4\}=
3\La^{\vt \mu}F_{\mu\nu}\xi^{\nu}+\xi^{\rho}\xi^{\mu}\dif_{\rho} F_{\mu\nu}\pi^{\nu}.
\e{505}
 If this is regarded as a
new constraint, then more new constraints are generated by their Poisson
brackets with the total Hamiltonian so that at the end no degree
of freedom remains. The generalized free constraints satisfy a Lie algebra apart from the 
  anomalous Poisson bracket relations \rl{505} and
  \be
 &&2\{\chi^\vt_1,\chi^\vt_3\}=3\La^\vt_\mu F^{\mu\nu}\pi_\nu-\pi^\rho\pi^\mu\dif_\rho F_{\mu\nu}\xi^\nu.
 \e{5051}
 Due to the Jacobi identity
 \be
 &&\{\chi^\vt_1,\chi^\vt_3\}=\{\{\chi^\vt_1, \chi^\vt_4\}, \chi^\vt_2\}
 \e{5052}
 the vanishing of \rl{505} implies the vanishing of \rl{5051} resulting in a closed Lie algebra of constraints identical to the one in section 3.
 On the other hand
 the only possibility for \rl{505} to vanish seems  to be $F_{\mu\nu}=0$ in which case the algebra of constraints of course is identical to the one for the free model.

 \subsubsection{Interaction with a vector field in the pseudoclassical case (index $\Lsa$)}
 When we add the minimal coupling term $A_{\mu}\dox^{\mu}$ to the
 pseudoclassical Lagrangian \rl{417} we find the Lagrangian
 \be
&L^\sa=&\sqrt{\biggl({d\over d\tau}\bigl(e\dox\bigr)\biggr)^2-2i\la_7\psi\cdot{d\over d\tau}\bigl(e\dox\bigr)}+{1\over e}\Xi+\nn\\&&+{i\over 2}\psi\cdot\dot{\psi}-{i\over 2}\tet\dot{\tet}-i\la_7\tet+A_{\mu}\dox^{\mu},
\e{5053}
and again we find that $p_{\mu}$ is
 replaced by $\La^\sa_{\mu}\equiv\La^\vt_{\mu}$ in \rl{5011}. The elementary
 Hamiltonian is  therefore also here given by \rl{502},   $H^\sa\equiv H^\vt$. The
 primary constraints are unaltered by the minimal coupling, \ie $\chi_2^\sa=\chi_2^\st=\chi_2$, $\chi_5^\sa=\chi_5^\st=\chi_5$, and $\chi_7^\sa=\chi_7^\st$ in \rl{409} (we ignore the fact that $\chi_7^\st$ is strictly speaking not a primary constraint).  The total
 Hamiltonian is the same as in the free case except for the different
 elementary Hamiltonian here ($H^\sa=H^\vt\neq H^\st$). Two secondary constraints are $\chi_3^\sa=\chi_3^\vt$ and $\chi_4^\sa=\chi_4^\vt$. Since
 \be
 &&\dot{\chi}^\sa_7=\{\chi_7^\sa, H^\sa\}=-\La^\sa\cdot\psi,
 \e{506}
 the condition $\dot{\chi}_7^{\sa}=0$ yields the secondary constraint 
 \be
 &&\chi^\sa_6\equiv\La^\sa\cdot\psi
 \e{507}
 in accordance with minimal coupling.
However, here we obtain
 \be
 &&\{\chi^\sa_6, H^\sa\}=\psi^{\mu}F_{\mu\nu}(x)\xi^{\nu},
 \e{508}
 which has to vanish from the consistency condition $\dot{\chi}^\sa_6=0$.
If it is interpreted as a new constraint then we  have a further proliferation of constraints from their consistency conditions. The field strength $F_{\mu\nu}$ has to be restricted. Apart from \rl{508} the algebra of constraints have the anomalous Poisson brackets \rl{505} and \rl{5051}, and ($\chi_1^\sa=\chi_1^\vt$ in \rl{5041})
 \be
 &&2\{\chi^\sa_1, \chi^\sa_7\}=\{\chi^\sa_6, \chi^\sa_3\}=\{\chi^\sa_2, \{\chi^\sa_4, \chi^\sa_6\}\}=-\pi^{\mu}F_{\mu\nu}\psi^{\nu},\nn\\
 &&2\{\chi^\sa_1, \chi^\sa_6\}=2\La^{\sa \mu}F_{\mu\nu}\psi^{\nu}+\dif_{\rho}F_{\mu\nu}\xi^{\mu}\pi^{\nu}\psi^{\rho},\nn\\
 &&\;\;\{\chi^\sa_6, \chi^\sa_6\}=-2i\chi^\sa_1+i\xi^{\mu}F_{\mu\nu}\pi^{\nu}+F_{\mu\nu}(x)\psi^{\mu}\psi^{\nu}.
 \e{509}
In order to have the same algebra as in the free case we have to impose $F_{\mu\nu}=0$.

\subsection{Gravitational interaction (index $\Lgt$)}
There is a  natural way to introduce gravitational
interaction in the model. If spacetime is Riemannian and curved,
the action must be independent of the choice of spacetime
coordinates and reduce to the free action for flat spacetime.
This leads us to the Lagrangian
\be
&&L^\gt=\sqrt{(\dot{e}\dot{x}^{\mu}+eD\dot{x}^{\mu})g_{\mu\nu}
(\dot{e}\dot{x}^{\nu}+eD\dot{x}^\nu)} + {1\over e}\Xi,\nn\\
&&D \dox^\mu
\equiv\ddot{x}^\mu+\dot{x}^\nu \Ga^\mu_{\nu\la}(x)\dot{x}^\la,
\e{510}
where $g_{\mu\nu}(x)$ is the metric tensor which serves as an external field, and
$\Ga^{\mu}_{\al\beta}(x)$  the Christoffel symbol:
\be
&&\Ga^{\mu}_{\al\beta}=\half g^{\mu\nu}(\dif_{\al}g_{\beta\nu}+\dif_{\beta}g_{\al\nu}-\dif_{\nu}g_{\al\beta}).
\e{511}
The Lagrangian \rl{510} is   reparametrization invariant both in $\tau$ and $x^{\mu}$.

Following Ostrodgradski's method with $\xi^{\mu}=\dox^{\mu}$ we are led to the following expressions for the conjugate momenta to $\xi^{\mu}$ and $e$
\be
&&\pi_{\mu}=\frac{eg_{\mu\nu}(\dot{e}\xi^{\nu}+eD{\xi}^{\nu})}{\sqrt{(\dot{e}\xi^{\la}+eD{\xi}^{\la})g_{\la\kappa}(\dot{e}\xi^{\kappa}+eD{\xi}^{\kappa})}},
\e{512}
\be
&&\omega=\frac{\xi^{\mu}g_{\mu\nu}(\dot{e}\xi^{\nu}+eD {\xi}^{\nu})}{\sqrt{(\dot{e}\xi^{\la}+eD {\xi}^{\la})g_{\la\kappa}(\dot{e}\xi^{\kappa}+eD {\xi}^{\kappa})}},
\\
&&D\xi^\mu
\equiv\dxi^\mu+\xi^\nu \Ga^\mu_{\nu\la}(x)\xi^\la.\nn
\e{513}
We have therefore the primary constraints
\be
&&2\chi^\gt_2\equiv\pi_{\mu}g^{\mu\nu}(x)\pi_{\nu}-e^2,\qquad\chi^\gt_5\equiv\pi_{\mu}\xi^{\mu}-\om e,
\e{514}
and the Hamiltonian
\be
&&H^\gt=\La^\gt_{\mu}\xi^{\mu}-{1\over e}\Xi,
\e{515}
where
\be
&&\La^\gt_{\mu}\equiv p_{\mu}-\pi_{\nu}\Ga^{\nu}_{\mu\la}(x)\xi^{\la}.
\e{516}
In the calculations of secondary constraints we notice that the Poisson bracket is the same as in subsection 3.1 except that curved indices are raised and lowered by the metric tensor $g_{\mu\nu}$. We notice also the following algebra
\be
&&\{\La^\gt_{\mu}, \La^\gt_{\nu}\}=\pi_{\rho}R^{\rho}_{\;\la\mu\nu}(x)\xi^\la,
\e{517}
where $R^{\rho}_{\;\la\mu\nu}$ is the Riemann tensor given by
\be
&&R^{\rho}_{\;\al\ga\beta}=\dif_\ga\Ga^{\rho}_{\al\beta}-\dif_{\beta}\Ga^{\rho}_{\al\ga}+\Ga^{\eta}_{\al\beta}\Ga^{\rho}_{\ga\eta}-\Ga^{\eta}_{\al\ga}\Ga^{\rho}_{\beta\eta}.
\e{518}
$\La^\gt_{\mu}$ acts like a covariant derivation according to the rule
\be
&&\{\pi_{\mu_1}\cdots\pi_{\mu_k}T^{\mu_1\cdots\mu_k}_{\nu_1\cdots\nu_l}(x)\xi^{\nu_1}\cdots\xi^{\nu_l}, \La^\gt_{\rho}\}=\pi_{\mu_1}\cdots\pi_{\mu_k}T^{\mu_1\cdots\mu_k}_{\nu_1\cdots\nu_l;\rho}(x)\xi^{\nu_1}\cdots\xi^{\nu_l},\nn\\
&&\nn\\
&&T^{\mu_1\cdots\mu_k}_{\nu_1\cdots\nu_l;\rho}(x)=\dif_{\rho}T^{\mu_1\cdots\mu_k}_{\nu_1\cdots\nu_l}(x)+\Ga^{\mu_1}_{\rho\ga}T^{\ga\mu_2\cdots\mu_k}_{\nu_1\cdots\nu_l}(x)+\cdots+\Ga^{\mu_k}_{\rho\ga}T^{\mu_1\cdots\mu_{k-1}\ga}_{\nu_1\cdots\nu_l}(x)\nn\\&&\qquad\qquad\quad\!-\Ga^{\ga}_{\rho\nu_1}T^{\mu_1\cdots\mu_k}_{\ga\nu_2\cdots\nu_l}(x)-\cdots -\Ga^{\ga}_{\rho\nu_l}T^{\ga\mu_2\cdots\mu_k}_{\nu_1\cdots\nu_{l-1}\ga}(x).
\e{5181}
 Vanishing time derivatives of $\chi^\gt_2$ and $\chi^\gt_5$, using the total Hamiltonian given by \rl{515} together with a linear combination of \rl{514},  requires  the following secondary  constraints ($g_{\mu\nu;\rho}=0$)
\be
&&\chi^\gt_3\equiv\pi_\rho g^{\rho\sigma}\La^\gt_\sigma,\qquad  \chi^\gt_4 \equiv H^\gt.
\e{519}
The conservation of $\chi^\gt_3$ requires in turn the tertiary constraint
\be
&&
2\chi^\gt_1\equiv
\La^\gt_\rho g^{\rho\sigma}\La^\gt_\sigma-\pi_{\rho}\pi_{\sigma}g^{\sigma\ga}R^\rho_{\;\al\ga\beta}\xi^\al\xi^\beta.
\e{5191}
 The consistency condition $\dot{\chi}^\gt_1=0$ leads, however, to new conditions. We have
\be
&&2\{\chi^\gt_1, \chi^\gt_4\}=4\La^\gt_\sigma g^{\sigma\ga}\pi_{\rho}R^\rho_{\;\al\ga\beta}(x)\xi^{\al}\xi^{\beta}
-\pi_{\rho}\pi_{\sigma}g^{\sigma\ga}R^\rho_{\;\al\ga\beta;\del}\xi^\al\xi^\beta\xi^\del,\nn\\
\e{520}
where $\dot{\chi}^\gt_1=0$ requires the right hand side to vanish. In order to avoid new constraints (particularly new  second class constraints), we have therefore to impose  restrictions on the metric tensor. If it is chosen such that \rl{520} vanishes then we also  get the same Poisson algebra as in the free case since the only further anomalous Poisson relation satisfies the following relations:
\be
&&2\{\chi^\gt_1, \chi^\gt_3\}=2\{\{\chi^\gt_1, \chi^\gt_4\}, \chi^\gt_2\}=\nn\\&&=8\La^\gt_\mu g^{\mu\beta}\pi_{\sigma}g^{\sigma\ga}\pi_{\rho}R^\rho_{\;\al\ga\beta}(x)\xi^{\al}
-3\pi_{\rho}\pi_{\sigma}g^{\sigma\ga}\pi_{\nu}g^{\nu\mu}R^\rho_{\;\al\ga\beta;\mu}\xi^\al\xi^\beta.
\e{5201}
In order to make \rl{520} vanish it seems, however, as we have to impose the restriction  $R^{\rho}_{\;\al\ga\beta}=0$ which leads to the free case in arbitrary coordinates. However, in subsection 5.3 below we consider less restrictive choices. 

 \subsubsection{Gravitational interaction in the pseudoclassical case\\ (index $\Lsgt$)}
In order to write the pseudoclassical model \rl{417} in an external gravitational field we have to make use of vierbein fields since the odd spin variables $\psi^{\mu}$ cannot have curved indices. In the vierbein formalism one usually denotes flat Minkowski indices by roman letters and as above curved indices by greek letters. The vierbein field, $V^a_\al(x)$, satisfies by definition the relation
\be
&&g_{\al\beta}(x)=V^a_{\al}(x)V^b_{\beta}(x)\eta_{ab}.
\e{522}
There is always an inverse, $V^{\al}_a(x)$, with the properties
\be
&&V^a_{\al}(x)V_a^{\beta}(x)=\del_{\al}^{\beta},\quad   V^a_{\al}(x)V^{\al}_b(x)=\del^a_b.
\e{5221}
In terms of the vierbein field
the pseudoclassical  Lagrangian  \rl{417} becomes in an external gravitational field
\be
&L^\sgt=&\sqrt{\bigl(\de\dox^{\mu}+eD\dox^{\mu}\bigr)g_{\mu\nu}\bigl(\de\dox^{\nu}+eD\dox^{\nu}\bigr)-2i\la_7\psi_aV^a_{\al}\bigl(\dot{e}\dox^{\al}+eD\dox^{\al}\bigr)}+\nn\\&&+{1\over e}\Xi+{i\over 2}\psi_aD{\psi}^a-{i\over 2}\tet\dot{\tet}-i\la_7\tet\equiv\nn\\
&&\equiv\sqrt{A^a\eta_{ab}A^b}+{1\over e}\Xi+{i\over 2}\psi_aD{\psi}^a-{i\over 2}\tet\dot{\tet}-i\la_7\tet, \nn\\&A^a=&\bigl(\de\dox^{\mu}+eD\dox^{\mu}\bigr)V^a_{\mu}-i\la_7\psi^a,
\e{521}
where $D\dox^{\mu}$ is defined in \rl{510}, and where
\be
&&D\psi^a\equiv\dot{\psi}^{a}+\om^a_{\;\;b\ga}\dox^{\ga}\psi^b,
\e{5222}
where $\om^a_{\;b\ga}$ is 
the spin connection. In order for it to yield a nonzero term in \rl{521} we must have $\om_{ab\ga}=-\om_{ba\ga}$ which also is its defining property.
Within the Hamiltonian formulation obtained from Ostrogradski's method we then find the elementary Hamiltonian
\be
&&H^\sgt=\La^\sgt_{\mu}\xi^{\mu}-{1\over e}\Xi, 
\e{5223}
where
\be
&&\La^\sgt_{\mu}\equiv p_{\mu}-\pi_{\sigma}\Ga^{\sigma}_{\mu\nu}\xi^{\nu}-{i\over 2}\psi^a\om_{ab\mu}\psi^b.
\e{5224}
We also notice the algebra
\be
\{\La^\sgt_{\mu},\La^\sgt_{\nu}\}=\pi_{\rho}R^{\rho}_{\;\la\mu\nu}\xi^{\la}+\frac{i}{2}\psi^a\psi^bR_{ab\mu\nu}
\e{5224b}
where
\be
R_{ab\mu\nu}=\dif_\mu\om_{ab\nu}-\dif_{\nu}\om_{ab\mu}+\om_{ac\mu}{\om^c}_{b\nu}-\om_{bc\mu}{\om^c}_{a\nu}=V^{\sigma}_aV^{\rho}_bR_{\sigma\rho\mu\nu}.
\e{5224c}
The Poisson bracket is the same as in subsection 4.1 except that curved indices are raised and lowered by the metric tensor $g_{\mu\nu}$ and flat indices by the Minkowski metric $\eta_{ab}$. 
$\La^\sgt_{\mu}$ generates in terms of this Poisson bracket generalized covariant derivatives on tensors with both curved and flat indices according to the rule ($m\leq4$)
\be
&&\{\pi_{\mu_1}\cdots\pi_{\mu_k}T^{\mu_1\cdots\mu_k}_{\nu_1\cdots\nu_l a_1\cdots a_m}(x)\xi^{\nu_1}\cdots\xi^{\nu_l}\psi^{a_1}\cdots\psi^{a_m}, \La^\sgt_{\rho}\}=\nn\\&&=\pi_{\mu_1}\cdots\pi_{\mu_k}T^{\mu_1\cdots\mu_k}_{\nu_1\cdots\nu_l a_1\cdots a_m;\rho}(x)\xi^{\nu_1}\cdots\xi^{\nu_l}\psi^{a_1}\cdots\psi^{a_m},\nn\\&&\nn\\
&&T^{\mu_1\cdots\mu_k}_{\nu_1\cdots\nu_la_1\cdots a_m;\rho}(x)=\dif_{\rho}T^{\mu_1\cdots\mu_k}_{\nu_1\cdots\nu_la_1\cdots a_m}(x)+\Ga^{\mu_1}_{\rho\ga}T^{\ga\mu_2\cdots\mu_k}_{\nu_1\cdots\nu_la_1\cdots a_m}(x)+\cdots\nn\\&&\qquad\qquad\quad\!-\Ga^{\ga}_{\rho\nu_1}T^{\mu_1\cdots\mu_k}_{\ga\nu_2\cdots\nu_la_1\cdots a_m}(x)-\cdots +\om_{a_1b\rho}\eta^{bc}T^{\mu_1\cdots\mu_k}_{\nu_1\cdots\nu_{l}ca_2\cdots a_m}(x)+\cdots.\nn\\
\e{5225}
Tensors with flat indices are directly related to tensors with curved indices by an appropriate multiplication of vierbein fields. The consistency condition for this is $V^\al_{a;\mu}=0$ which also yields $g^{\al\beta}_{\;\;\;;\mu}=0$. This condition determines the spin connection. We find
\be
&&\om_{ab\mu}(x)=V^{\nu}_a\bigl(\dif_{\mu}V_{b\nu}-\Ga^{\rho}_{\mu\nu}V_{b\rho}\bigr).
\e{5226}
Ostrogradski's method yields apart from the Hamiltonian \rl{5223} and the constraints $\chi^\sgt_2=\chi^\gt_2$ and $\chi^\sgt_5=\chi^\gt_5$ in
\rl{514}, ${\chi}^\sgt_7=0$ where 
\be
&&\chi^\sgt_7\equiv\pi_{\al}V^{\al}_a\psi^a+e\tet.
\e{523}
 Requiring $\chi^\sgt_2$, $\chi^\sgt_5$ and $\chi^\sgt_7$ to be constants of
 motion (Dirac's consistency condition) leads to the secondary
 constraints $\chi^\sgt_4=H^\sgt$ given in \rl{5223}, $\chi^\sgt_3=0$ and $\chi^\sgt_6=0$ where 
 \be
&&\chi^\sgt_3\equiv
\pi_{\mu}g^{\mu\sigma}\La^\sgt_{\sigma},\qquad\chi^\sgt_6\equiv\La^\sgt_{\al}V^{\al}_a\psi^a.
\e{524}
At the tertiary level we have  ($\dot{\chi}^\sgt_3=0$)
\be
2\chi^\sgt_1\equiv\La^\sgt_{\al}g^{\al\beta}\La^\sgt_{\beta}-\pi_{\mu}g^{\mu\nu}\xi^{\la}(\pi_{\ga}R^{\ga}_{\;\rho\nu\la}\xi^{\rho}+\frac{i}{2}R_{ab\nu\la}\psi^a\psi^b)
\e{5242}
and
\be
&&\{\chi^\sgt_6, \chi^\sgt_4\}=R_1,\nn\\
&&R_1\equiv V^\mu_c\psi^c\xi^{\nu}(\pi_{\rho}R^\rho_{\;\la\mu\nu}\xi^\la+\frac{i}{2}R_{ab\mu\nu}\psi^a\psi^b).
\e{525}
$\dot{\chi}^\sgt_6=0$ requires the new condition
$R_1=0$. $\dot{\chi}^\sgt_1=0$ requires apart from the vanishing of (cf.\rl{520})
\be
&2\{\chi^\sgt_1,\chi^\sgt_4\}&\!\!\!\!=4\La^\sgt_{\al}\pi_{\rho}g^{\al\beta}R^{\rho}_{\;\mu\beta\nu}\xi^{\mu}\xi^{\nu}+\frac{3}{2}i\La^\sgt_{\al}g^{\al\beta}R_{ab\beta\mu}\xi^{\mu}\psi^a\psi^b\nn\\
&&-\pi_{\mu}g^{\mu\nu}\xi^{\al}\xi^{\la}(\pi_{\ga}\xi^{\beta}R^{\ga}_{\;\beta\nu\al;\la}+\frac{i}{2}R_{ab\nu\al;\la}\psi^a\psi^b)
\e{5252}
also $R_2=0$ where
\be
&&\{\chi^\sgt_1, \chi^\sgt_7\}=R_2,\nn\\
&&R_2=\frac{1}{2}\pi_{\rho}\pi_{\sigma}g^{\sigma\ga}R^\rho_{\;\al\beta\ga}\xi^\al V^{\beta}_a\psi^a.
\e{5251}
In addition we  notice that (cf.\rl{5201})
\be
&&2\{\chi^\sgt_1,\chi^\sgt_3\}=2\{\{\chi^\sgt_1,\chi^\sgt_4\},\chi^\sgt_2\}=\nn\\
&&=8\La^\sgt_{\al}\pi_{\rho}\pi_{\sigma}g^{\sigma\nu}g^{\al\beta}R^{\rho}_{\;\mu\beta\nu}\xi^{\mu}+\frac{3}{2}i\La^\sgt_{\al}\pi_{\sigma}g^{\sigma\mu}g^{\al\beta}R_{ab\beta\mu}\psi^a\psi^b\nn\\
&&\quad-\pi_{\mu}\pi_{\sigma}g^{\sigma\alpha}g^{\mu\nu}\xi^{\la}(3\pi_{\ga}\xi^{\beta}R^{\ga}_{\;\beta\nu\al;\la}+iR_{ab\nu\la;\al}\psi^a\psi^b).
\e{5253}
Furthermore, we have also the following anomalous Poisson brackets as compared to the free algebra in section 4:
\be
&&\{\chi^\sgt_6, \chi^\sgt_3\}=2R_2,\quad\{\chi^\sgt_6, \chi^\sgt_6\}=-2i\chi^\sgt_1+iR_3,\quad\{\chi^\sgt_6, \chi^\sgt_1\}=R_4,\nn\\
\e{526}
where
\be
&&R_3\equiv \pi_{\rho}\pi_\la g^{\la\ga}R^{\rho}_{\;\al\beta\ga}\xi^\al\xi^\beta-\frac{3}{2}i\pi_{\rho}R^{\rho}_{\;\al\beta\ga}\xi^\al V^\beta_aV^\ga_b\psi^a\psi^b,\nn\\
&&R_4\equiv \frac{1}{2}\pi_{\rho}R^\rho_{\;\al\beta\ga}\xi^\al
g^{\ga\sigma}\La^\sgt_{\sigma}V^\beta_a\psi^a+\half iV^{\beta}_c\psi^c\La^\sgt_{\mu}g^{\mu\nu}R_{ab\beta\nu}\psi^a\psi^b\nn\\
&&\qquad-\half\pi_\sigma g^{\sigma\ga}\xi^{\al}(\pi_{\rho}R^{\rho}_{\;\al\beta\ga;\mu}\xi^\beta+\frac{i}{2}R_{ab\ga\al;\mu}\psi^a\psi^b)V^{\mu}_c\psi^c.
\e{527}
The expressions $R_1$-$R_4$ and \rl{5252}, \rl{5253} represent deviations from a closed Poisson algebra. Their vanishing without additional constraints requires $R^{\rho}_{\;\al\beta\ga}=0$. Notice that the vanishing of $R_1$, $R_2$, and \rl{5252} implies the vanishing of $R_3$, $R_4$, and \rl{5253} since we have from the Jacobi identities \rl{5253} and
\be
&&R_3=-i\{R_1, \chi^\sgt_7\},\qquad R_4=\{R_2, \chi^\sgt_4\}-\{\{\chi^\sgt_1, \chi^\sgt_4\}, \chi^\sgt_7\}.
\e{528}

\subsection{Infinite spin particles in (anti)de Sitter spacetime (index $\LDS$)}
From the previous results it seems as if we only have consistent classical models in flat spacetime. However, an obvious  question is whether or not there are special curved spacetimes for which there are a finite number of constraints satisfying a closed Poisson bracket algebra.      In order to investigate this we consider here maximally symmetric spacetimes for which we have   ($K$ is a real constant which is positive for a de Sitter space, and negative for an anti-de Sitter space)
\be
&&R_{\mu\al\beta\ga}=K\bigl(g_{\al\beta}g_{\mu\ga}-g_{\al\ga}g_{\beta\mu}\bigr),\;\;\Rightarrow\;\; R_{\mu\al\beta\ga;\nu}=0.
\e{530}
For this choice \rl{517} reduces to ($\La^\DS_{\mu}=\La^\gt_{\mu}$ in (anti)de Sitter space)
\be
&&\{\La^{\DS}_\mu, \La^{\DS}_\nu\}=K\bigl(\pi_\nu\xi_\mu-\pi_\mu\xi_\nu\bigr).
\e{531}
Furthermore, eq.\rl{520} reduces to ($\chi^\DS_i=\chi^\gt_i$ in (anti)de Sitter space)
\be
&&\{\chi^{\DS}_1, \chi^{\DS}_4\}=2K\bigl(\pi_\al\xi^\al\chi^{\DS}_4-\xi^\al g_{\al\beta}\xi^\beta\chi^{\DS}_3+{\Xi\over e}\chi_5^{\DS}+\Xi\om\bigr).
\e{532}
Thus, for maximally symmetric spacetimes, in which \rl{530} is valid, Dirac's consistency conditions yield a finite number of constraints provided $\Xi=0$. Since their Poisson algebra closes the resulting theory is consistent.  However, even though the algebra of constraints closes for $\Xi=0$ it does not form a Lie algebra.
 It is the constraint $\chi^\DS_1$ that causes  problems. For the (anti)de Sitter metric in \rl{530} we have
 \be
&&2\chi^{\DS}_1=\La^{\DS}_\rho g^{\rho\sigma}\La^{\DS}_{\sigma}+K\bigl(2\xi^\al g_{\al\beta}\xi^\beta\chi^{\DS}_2-(\chi_5^{\DS})^2-2\om e\chi_5^{\DS}+e^2(\xi^\al g_{\al\beta}\xi^\beta-\om^2)\bigr).\nn\\
\e{533}
This implies that the consistency condition $\dot{\chi}^{\DS}_3=0$ allows us to choose a more elementary form for $\chi^{\DS}_1$  like
\be
&&2\chi^{\prime \DS}_1\equiv\La^{\DS}_\rho g^{\rho\sigma}\La^{\DS}_{\sigma}+Ke^2(\xi^\al g_{\al\beta}\xi^\beta-\om^2).
\e{534}
However, even if $\chi^{\DS}_1$ is replaced by $\chi^{\prime \DS}_1$ the constraints still do not form a Lie algebra although they close. The pseudoclassical case below provides further insights.

\subsubsection{The pseudoclassical model in (anti)de Sitter space\\ (index $\LsDS$)}
In the pseudoclassical case the consistency condition
$\dot{{\chi}}^{\sDS}_1=0$ not only requires the vanishing of \rl{5252}
but also of $R_2$ in \rl{5251}. ($\chi^\sDS_i=\chi^\sgt_i$ in (anti)de
Sitter spacetime.) With the (anti)de Sitter metric \rl{530} $R_2$ becomes
\be
&&R_2=K\bigl(\pi_{\al}\xi^{\al}\chi^{\sDS}_7-2\xi_{\mu}V^{\mu}_a\psi^a\chi^{\sDS}_2-\tet e\chi_5^{\sDS}\bigr)-Ke^2\bigl(\tet\om+\xi_{\mu}V^{\mu}_a\psi^a\bigr)\nn\\
\e{535}
and
\be
&&\{\chi^{\sDS}_1, \chi^{\sDS}_4\}=2K\bigl(\frac{3}{4}i\psi^aV_a^{\mu}g_{\mu\nu}\xi^{\nu}\chi^{\sDS}_6-\xi^\al g_{\al\beta}\xi^\beta\chi^{\sDS}_3+\pi_\al\xi^\al\chi^{\sDS}_4+{\Xi\over e}\chi_5^{\sDS}+\Xi\om\bigr).\nn\\
\e{5352}
Hence, $R_2=0$ yields a new constraint given by 
\be
&&\chi^{\sDS}_9\equiv\tet\om+\xi_{\mu}V^{\mu}_a\psi^a.
\e{536}
The consistency condition $\dot{\chi}^{\sDS}_6=0$ requires the vanishing of $R_1$ in \rl{525}. In (anti)de Sitter spacetime we have
\be
&&R_1=K\bigl(-\xi^{\mu}g_{\mu\nu}\xi^{\nu}\chi^{\sDS}_7+\xi_{\mu}V^{\mu}_a\psi^a\chi_5^{\sDS}+e\om\chi^{\sDS}_9\bigr)+K\tet e\bigl(\xi^{\mu}g_{\mu\nu}\xi^{\nu}-\om^2\bigr).\nn\\
\e{537}
Hence, $R_1=0$ yields one more new constraint given by
\be
&&2\chi^{\sDS}_8\equiv \xi^{\mu}g_{\mu\nu}\xi^{\nu}-\om^2.
\e{538}
(General $\chi^\sgt_9$ and $\chi^\sgt_8$  may be defined, but not derived within the general formalism in subsection 5.2.1.)
The theory is consistent since the Poisson algebra of the constraints
closes.  However, their algebra does not form a Lie algebra. In the
pseudoclassical case we have for the (anti)de Sitter metric \rl{530} (cf.(\rl{533})
\be
&&2\chi^{\sDS}_1=\La^{\sDS}_\rho g^{\rho\sigma}\La^{\sDS}_{\sigma}+K\bigl(2\xi^\al g_{\al\beta}\xi^\beta\chi^{\sDS}_2-(\chi_5^{\sDS})^2-2\om e\chi_5^{\sDS}-iV_a^{\al}\psi^a\xi_{\al}\chi_7^{\sDS}\nn\\&&\qquad+e^2(\xi^\al g_{\al\beta}\xi^\beta-\om^2)+iV_a^{\al}\psi^a\xi_{\al}e\theta\bigr).
\e{539}
Even if we define $\chi^{\sDS}_1$ by $2\chi^{\prime\prime \sDS}_1\equiv \La^{\sDS}_{\mu}g^{\mu\nu}\La^{\sDS}_{\nu}$ (see \rl{533})  we still have no Lie algebra although the constraint algebra closes. The nontrivial part comes from  the internal algebra of $\chi^{\prime\prime\sDS}_1$, $\chi^{\sDS}_3$ and $\chi^{\sDS}_4$.

\setcounter{equation}{0}
\section{The standard massless representation}
For $\Xi=0$ we have to solve
\be
&&p^2|phys\hb=0,\quad w^2|phys\hb=0,
\e{601}
which according to the analysis in section 2 leads to the elementary set \rl{12}. To construct the corresponding classical theory we start from the Hamiltonian
\be
&&H=\la_1\chi_1+\la_3\chi_3+\la_4\chi_4,
\e{602}
where
\be
&&\chi_1\equiv \half p^2,\quad \chi_3\equiv p\cdot\pi,\quad \chi_4\equiv p\cdot\xi,
\e{603}
which satisfy a  Poisson algebra which is a nilpotent Lie algebra.
The Lagrangian is then according to the analysis in section 3 either given by \rl{304} with $\Xi=0$, $\la_2=0$ and $\la_3\lra\la_4$, or \rl{305} with $\Xi=0$, $\la_2=0$. In both cases we are led to the unique Lagrangian
\be
&&L=-{\la_1\over 2\la_3^2}\dxi^2+{1\over\la_3}(\dox-\la_4\xi)\cdot\dxi.
\e{604}
The constraints in the configuration space are 
\be
&&\dxi^2=0,\qquad\xi\cdot\dxi=0,\qquad\dox\cdot\dxi=0.
\e{605}
In this case it is not possible to eliminate $\xi^\mu$ from $L$ as in section 3 since $\la_2=0$ here.
The equations of motion imply, however, that $\xi^2$ and $\pi^2$ are constants of motions. We may therefore consistently add further constraints like $\chi_2=0$ where $\chi_2$ \eg is given by 
\be
&&\chi_2\equiv\half\bigl(\pi^2-1\bigr).
\e{606}
For this choice we then arrive at the Lagrangian \rl{311} with $\Xi=0$. The resulting $\Xi=0$ model is then just a particular choice of the general $\Xi\neq0$ model considered before. In the $\Xi=0$ case we may also consistently impose the further constraint $\chi_8=0$, where (cf.\@\rl{538})
\be
&&\chi_8\equiv \half\bigl(\xi^2-\om^2\bigr),
\e{607}
to the constraints following from \rl{311} in the Hamiltonian form. The resulting Lie algebra of the constraints is then a semi-direct sum of the nilpotent algebra of \rl{603} and $sl(2,R)$. Explicitly it is given in \rl{3051}, \rl{324} and (the nonzero part)
\be
&&\{\chi_8, \chi_2\}=\chi_5,\quad\{\chi_8, \chi_3\}=\chi_4,\quad\{\chi_8, \chi_5\}=2\chi_8.
\e{6071}

A Lagrangian for this model may be constructed following the procedure of section 3. Starting from the Hamiltonian 
\be
&&H=\la_1\chi_1+\la_2\chi_2+\la_3\chi_3+\la_4\chi_4+\la_5\chi_5+\la_8\chi_8
\e{6072}
 we find for $\la_1=\la_5=0$ ($\chi_1$ and $\chi_5$ are generated by the consistency conditions)
\be
&&L=-{\la_2\over 2\la_3^2}\bigl(\dox-\la_4\xi\bigr)^2+{1\over\la_3}\bigl(\dox-\la_4\xi\bigr)\cdot\dxi-\half\la_8\xi^2-{1\over 2\la_8}\dot{e}^2+\half\la_2 e^2.\nn\\
\e{608}
(Even in \rl{604} we may set $\la_1=0$ since $\chi_1$ is generated by the consistency condition $\dot{\chi}_4=0$.) Even if we now may eliminate $\xi^{\mu}$ from the equations of motion of $\xi^{\mu}$ choosing $\la_3=\la_4$, which is allowed, we do not obtain an equivalent geometrical higher order theory. The reason is that $1/\la_4$ becomes dynamical but different from $e$ which we in section 3 defined by $e=1/\la_4$. (If we have no $e$'s to start with we do not get a closed algebra for the constraints.) We do not know whether or not there exist a
 geometrical higher order Lagrangian  for this extended model. 
 
 In the pseudoclassical version we may apart from \rl{607} also add the constraint
 \be
 &&\chi_9\equiv\xi_{\mu}\psi^{\mu}+\om\tet.
 \e{609}
 The resulting Lie algebra is given by \rl{3051}, \rl{324}, \rl{6071}, \rl{410}, \rl{427} and (the nonzero part)
 \be
 &&\{\chi_9, \chi_2\}=\chi_7,\quad\{\chi_9, \chi_3\}=\chi_6,\quad\,\,\,\,\{\chi_9, \chi_5\}=\chi_9,\nn\\
 &&\{\chi_9, \chi_6\}=i\chi_4,\quad\{\chi_9, \chi_7\}=i\chi_5,\quad\{\chi_9, \chi_9\}=-2i\chi_8.
 \e{610}
 This is exactly the algebra obtained from the geometrical higher order Lagrangian in (anti)de Sitter spacetime given in subsection 5.3.1 (apart from the internal algebra of $\chi_1$, $\chi_3$, and $\chi_4$) provided we choose $2\chi_1=\La_{\mu}g^{\mu\nu}\La_{\nu}$.

\setcounter{equation}{0}
 \section{Quantization}
 Our derivations of the free models were made from the quantum treatments in \cite{Bargmann:1948gr,Wigner:1963in} and their generalizations. In these derivations  in section 2 and in the beginning of section 4 we derived quantum equations which are nothing else but a Dirac quantization of the considered models. A Dirac quantization is characterized by the following properties: The constraints $\chi_i$ are turned into operators $\hat{\chi}_i$ which are hermitian and satisfy a commutator Lie algebra,
 \be
 &&[\hat{\chi}_i, \hat{\chi}_j]=iC_{ijk}\hat{\chi}_k, \quad\forall i,j,k
 \e{701}
 where $C_{ijk}$ are real constants. The physical states are then consistently defined by the equations
 \be
 &&\hat{\chi}_i|phys\hb=0, \quad\forall i.
 \e{702}
 In a wave function representation \rl{702} is turned into wave equations. The $O(\Xi)$-representation considered in section 2 were in \cite{Bargmann:1948gr,Wigner:1963in}   given as wave equations of the type
 \be
 &&D_i\Phi(x,\xi)=0, \quad\forall i,
 \e{703}
 where $D_i$ is a differential operator representation of $\hat{\chi}_i$. $\Phi(x,\xi)$ is here just a scalar wave function of a bilocal type. For the $0'(\Xi)$-representation it is turned into a spinor, $\Phi_{\al}(x,\xi)$,  satisfying the Dirac equation ${\slash\hspace{ -2.2mm}\dif}\Phi(x,\xi)=0$ \cite{Bargmann:1948gr}. 
 
 Now we believe that a Dirac quantization of our free models is inconsistent. As arguments for this belief we give here some negative features of the procedure \rl{702},\rl{703}. We expect the quantization of the models to yield equations for higher spin fields. Therefore we should be able to derive  
 equations for tensor fields which are the standard form for a covariant description of higher spin fields \cite{Fierz:1939su,*Fierz:1939on}. One natural way to obtain such equations from \rl{703} is to Taylor expand the  bilocal field  in terms of $\xi$. Unfortunately, the resulting equations leave no non-zero solutions at all.

 A second approach would be to replace $\pi$ and $\xi$ by the oscillator $a$ defined by
 \be
 &&a^{\mu}\equiv{1\over\sqrt{2}}\bigl(\xi^{\mu}+i\pi^{\mu}\bigr)\quad\Rightarrow\quad[a^{\mu}, a^{\nu\dag }]=\eta^{\mu\nu}.
 \e{704}
 The constraint operators in \rl{9} for $F=1$ become then
\be
 &&\hat\chi_1\equiv\half p^2,\qquad\qquad\qquad\quad\hat\chi_2\equiv-\half\bigl((a-a\dagg)^2+1\bigr),\nn\\
 &&\hat\chi_3\equiv{i\over\sqrt{2}}\bigl(p\cdot a\dagg-p\cdot a\bigr),\quad\hat\chi_4\equiv{1\over\sqrt{2}}\bigl(p\cdot a+p\cdot a\dagg\bigr)-\Xi.
 \e{705}
 They may be obtained by the analysis in section 2 starting from
 \be
 &&s^{\mu\nu}=-i\bigl(a^{\mu\dag}a^{\nu}-a^{\nu\dag}a^{\mu}\bigr),
 \e{706}
  which is identical to the expression in \rl{2} using \rl{704}. The equations \rl{8} with the Fock ansatz
  \be
  &&|\psi\hb=\sum_{k=0}^\infty{1\over k!}A_{\mu_1\mu_2\cdots\mu_k}(x)|0\hb^{\mu_1\mu_2\cdots\mu_k},\nn\\
  &&|0\hb^{\mu_1\mu_2\cdots\mu_k}\equiv a^{\mu_1\dag}a^{\mu_2\dag}\cdots a^{\mu_k\dag}|0\hb,\quad p_{\mu}|0\hb=0,
  \e{707}
  yield equations for the $A$-fields which leave no non-zero solutions. In fact, these equations are the same as those obtained by a Taylor expansion of $\Phi(x,\xi)$ in $\xi$ above. Negative results were also  obtained    in  \cite{Abbott:1976ma,*Hirata:1977qu} where  the same equations were treated noncovariantly.
  
  The models considered in this paper are gauge theories and the general framework to quantize gauge theories is the BRST-quantization. This procedure is based on the use of an odd BRST-charge $Q$ to project out the physical states by the single condition $Q|phys\hb=0$. For the models under considerations in this paper we believe that BRST quantization is the correct quantization procedure. Furthermore, we believe that such a BRST quantization is inconsistent with the Dirac quantization used in sections 2 and 4 and above.  Now a
 BRST-approach requires states with finite inner products. This implies among other things that the derivations of representations in section 2 should be performed in a weak sense, \ie we should solve conditions like $\vb phys|(w^2-\Xi^2)|phys\hb=0$.  Such conditions lead naturally to a kind of  Gupta-Bleuler quantization.  Below we try to simulate a correct treatment in terms of such a Gupta-Bleuler quantization  in order to avoid the complexity of a fully fledged BRST treatment. We keep then the Dirac condition for $\hat\chi_1$ which leads to Klein-Gordon like equations as well as for $\hat\chi_6$ in section 4 which yields Dirac like equations. Their proper treatments within a BRST frame is known.
 
 \subsection{Gupta-Bleuler quantization}
 Gupta-Bleuler quantization is characterized by  conditions of the type
 \be
 &&G_r|phys\hb=0,\quad \forall r,
 \e{708}
 where $G_r$ are operators which not need to be hermitian but which must satisfy the following properties,
 \be
 &&[G_r, G_s]=iC'_{rst}G_t,
 \e{709}
where $C'_{rst}$ are constants not necessarily real, and where
\be
&&\vb phys|{\hat\chi}_i|phys\hb=0,\quad \forall i,
\e{710}
are implied by \rl{708}. Of course, \rl{708} includes the Dirac quantization. However, the number of $G_r$-operators are usually less than the number of ${\hat\chi}_i$'s. For the physical states  we use the following general Fock-like ansatz
\be
  &&|\psi\hb=\sum_{n=-\infty}^{\infty}\sum_{k=1}^\infty{1\over k!}A^{(n)}_{\mu_1\mu_2\cdots\mu_k}(x)|0,n\hb^{\mu_1\mu_2\cdots\mu_k}+\!\!\!\sum_{n=-\infty}^{\infty}\phi^{(n)}(x)|0,n\hb,\nn\\
  &&|0,n\hb^{\mu_1\mu_2\cdots\mu_k}\equiv a^{\mu_1\dag}a^{\mu_2\dag}\cdots a^{\mu_k\dag}|0,n\hb, \quad |0,n\hb\equiv|0\hb_p |0\hb |n\hb, \quad a^{\mu}|0\hb=0\nn\\
&& p_{\mu}|0\hb_p=0, \quad  |n\hb\equiv e^n|0\hb_{\omega}, \quad\omega |0\hb_{\omega}=0
  \e{711}
where $a^{\mu}$ is defined in \rl{704} and where $e$ is the inverse einbein introduced in section 3.1 with $[e,\omega]=i$ and $e\neq0$. It should be noted that neither $|0\hb_p$ nor $|n\hb$ are inner product states. ($|n\hb$-states are \eg  discussed in \cite{Marnelius:1993ge}.) Our equations are therefore a bit heuristic.

The wave function representation of the ansatz \rl{711} is given by
\be
&&\psi(x,e)\equiv\vb x,e|\psi\hb=\sum_n \phi^{(n)}(x) e^n,\nn\\
&&\psi_{\mu_1\cdots\mu_k}(x,e)\equiv \;_{\mu_1\cdots\mu_k}\vb x,e|\psi\hb=\sum_n A^{(n)}_{\mu_1\cdots\mu_k}(x) e^n,\quad k\geq1,
\e{7111}
where $e\neq0$ and
\be
&&_{\mu_1\cdots\mu_k}\vb x,e|\equiv \vb x|\vb e|\vb0|a_{\mu_1}\cdots a_{\mu_k}.
\e{7112}
Here the inverse einbein variable $e$ acts like an extra dimension. The true spacetime wave function may be defined by the gauge fixed expression
\be
&&A_{\mu_1\cdots\mu_k}(x) \equiv\psi_{\mu_1\cdots\mu_k}(x,e_0) =\sum_n A^{(n)}_{\mu_1\cdots\mu_k}(x) e_0^n,
\e{7113}
where $e_0$ is a fixed value of the $e$-variable. Or, possibly, it could be defined to be some other weighted sum of the $A^{(n)}$-fields. The proper interpretation remains to be investigated.

 \subsection{Quantization of the free classical model}
A Gupta-Bleuler quantization for the free theory considered in section 3.1 may be performed by means of the following constraint operators:
\be
&&G_0\equiv 2\hat\chi_1, \nn\\
&&G_1\equiv \frac{1}{\sqrt 2}(\hat\chi_4+i\hat\chi_3), \nn\\
&&G_2\equiv 2 \hat\chi_2-i\hat\chi_5
\e{712}
which satisfy the Lie algebra ($G_0$ commutes with $G_{1,2}$)
\be 
&&[G_1,G_2]=G_1.
\e{713}
The constraint operators $\hat\chi_1,\dots, \hat\chi_5$ correspond to those  given in section 3.1 (see also appendix A.1). They are explicitly given by \rl{705} and
\be
&&\hat\chi_5\equiv{i\over2}\bigl(a^{\dag 2}-a^2\bigr)-\half\bigr(\om e+e\om\bigr),
\e{7131}
where $\xi$ and $\pi$ are given in terms of the oscillator $a$ defined in \rl{704}. The expressions \rl{712} are therefore
 \be
&&G_0= p^2, \nn\\
&&G_1= p\cdot a -\frac{\Xi}{\sqrt{2} e},\nn\\
&&G_2= -a^2+\frac{1}{2} \biggl(aa^\dag+a^\dag a-2e^2+i(\omega e +e\omega)\biggr).
\e{714} 
These constraint operators can be derived by the analysis in section 2 starting from \rl{706} using  weak conditions.  In fact, we have
\be
&&\vb ph|e^2\bigl(w^2-\Xi^2\bigr)e^2|ph\hb=\biggl|G_{0,1}|ph\hb=0\biggr|=\nn\\
&&={\Xi^2\over2}\biggl(\vb ph|G_2^{\dag}e^2|ph\hb+\vb ph|e^2G_2|ph\hb\biggr)=0.
\e{715}
The peculiar factors $e^2$ are required by the form of $G_2$ which in turn is chosen to satisfy the simple algebra \rl{713}.
 The constraint operators in \rl{712},\rl{714} on the Fock ansatz \rl{711} 
 yield now non-zero solutions for the $A$- and $\phi$-fields. The $G_0|\psi\hb=0$ condition simply yields the Klein-Gordon equations
\be
&&\dif^2 \Phi^{(n)}(x)=0, \quad \dif^2 A^{(n)}_{\mu_1,\dots, \mu_k}=0,
\e{716}
whereas $G_1|\psi\hb=0$ yields
\be
&&i\dif^{\nu}A_{\nu}^{(n)}(x)+\frac{\Xi}{\sqrt{2}}\phi^{(n+1)}(x)=0, \nn\\
&&i\dif^{\nu}A^{(n)}_{\nu\mu_1\dots\mu_k}(x)+\frac{\Xi}{\sqrt{2}}A^{(n+1)}_{\mu_1\dots\mu_k}(x)=0,\quad k\geq1,
\e{717}
and $G_2|\psi\hb=0$ yields finally
\be
&&\biggl(n+{5\over2}\biggr)\phi^{(n)}(x)-A^{(n)\nu}_{\nu}(x)-\phi^{(n-2)}(x)=0,\nn\\
&&\biggl(n+k+{5\over2}\biggr)A_{\mu_1\dots\mu_k}^{(n)}(x)-A^{(n)\nu}_{\nu\mu_1\dots\mu_k}(x)-A^{(n-2)}_{\mu_1\dots\mu_k}(x)=0,\quad k\geq1.\nn\\
\e{718}

It remains to investigate what these relations actually imply for the true spacetime wavefunctions (see subsection 7.1).

 \subsection{Quantization of the simple free $\Xi=0$ model} 
The constraints for the simple free $\Xi=0$ model considered in section 6 (see also appendix A.4) are here combined into the Gupta-Bleuler operators 
\be
&&G_0\equiv 2\hat\chi_1=p^2, \nn\\
&&G_1\equiv \frac{1}{\sqrt 2}(\hat\chi_4+i\hat\chi_3)=p\cdot a.
\e{720}
As expected the constraints $G_0|\psi\hb=0$ and $G_1|\psi\hb=0$ now yield massless Klein-Gordon equations and Lorentz like conditions
\be
&&\dif^2 \phi^{(n)}(x)=0, \quad \dif^2 A^{(n)}_{\mu_1\dots \mu_k}=0,\nn\\
&&\dif^{\nu}A^{(n)}_{\nu\mu_1\dots\mu_k}=0,\quad k\geq1.
\e{721}

These equations imply the same equations for the true spacetime functions whatever way they are defined.

 \subsection{Quantization of the extended free $\Xi=0$ model} 
 Let us consider the extended free $\Xi=0$ representation consisting of the constraints $\hat\chi_1,\dots, \hat\chi_5$ with the additional constraint $\hat\chi_8$. This model was studied in section 6 (see also appendix A.5).  The Gupta-Bleuler constraint operators are here chosen to be
 \be
 &&G_0\equiv2\hat\chi_1=p^2,\nn\\
 &&G_1 \equiv\frac{1}{\sqrt 2}(\hat\chi_4+i\hat\chi_3)=p\cdot a,\nn\\
 &&G_3 \equiv\hat\chi_2+\hat\chi_8=a^{\dag}a+2-\frac{1}{2}(e^2+\omega^2),\nn\\
 &&G_4 \equiv\frac{1}{2}(\hat\chi_8-\hat\chi_2+i\hat\chi_5)=a^2+\frac{1}{2}(e^2-\omega^2)-\frac{i}{2}(\omega e+e\omega),
 \e{722} 
satisfying the algebra
\be
&&[G_1,G_3]=G_1, \quad [G_3,G_4]=-2G_4.
\e{723} 
Using the Fock ansatz \rl{711}, $G_0|\psi\hb=0$ and $G_1|\psi\hb=0$  yield the Klein-Gordon field equations and the Lorentz conditions in \rl{721}.
Restrictions on the fields are obtained  by the conditions from $G_3|\psi\hb=0$ given by
\be
&&\!\!\!\!2\phi^{(n)}(x)-{1\over 2}\phi^{(n-2)}(x)+{1\over 2} (n+2)(n+1)\phi^{(n+2)}(x)=0,\nn\\
&&\!\!\!\!(k+2)A^{(n)}_{\mu_1\dots\mu_k}(x)-{1\over 2}A_{\mu_1\dots\mu_k}^{(n-2)}(x)+{1\over 2} (n+2)(n+1)A_{\mu_1\dots\mu_k}^{(n+2)}(x)=0,\quad k\geq1.\nn\\
\e{725}
 Furthermore, from $G_4|\psi\hb=0$ we find
 \be
&&A^{(n)\nu}_{\nu}(x)+{1\over 2}\phi^{(n-2)}(x)+{1\over 2} (n+2)(n+1)\phi^{(n+2)}(x)-(n+{1\over2})\phi^{(n)}(x)=0,\nn\\
&&A^{(n)\nu}_{\nu\mu_1\dots\mu_k}(x)+{1\over 2}A_{\mu_1\dots\mu_k}^{(n-2)}(x)+{1\over 2} (n+2)(n+1)A_{\mu_1\dots\mu_k}^{(n+2)}(x)-\nn\\
&&\quad-(n+{1\over2})A^{(n)}_{\mu_1\dots\mu_k}(x)=0,\quad k\geq1.
\e{726}

Again it remains to investigate the implications of these equations for the true spacetime wave functions.

 In general for representations with $p^2=0$ and $w^2=0$ we may construct a covariant helicity operator $\la$ from $w^{\mu}+\la p^{\mu}=0$. However, just in the particular case when the internal variable is a bosonic oscillator $a^{\mu}$ this is not possible (see section 5 in \cite{Marnelius:1990de}).
%
%
\subsection{Quantization of the free pseudoclassical model}
There are problems to quantize the main pseudoclassical model given in subsection 4.1 according to the Gupta-Bleuler scheme. The problem lies in the quantum constraint $\hat\chi^s_7$ which satisfies the relation $(\hat\chi^s_7)^2=\hat\chi^s_2$. We do not know how it should fit into a choice of Gupta-Bleuler constraints. We expect that a BRST treatment should solve this dilemma. However, without doing the appropriate analysis we are unable to guess a possible solution.

Without the $\hat\chi^s_7$-constraint the Gupta-Bleuler quantization is straight-forward. We may then choose the Gupta-Bleuler constraints as follows
\be
 &&G^s_0 \equiv 2\hat\chi^s_1= p^2,\nn\\
 &&G^s_1 \equiv \frac{1}{\sqrt 2}(\hat\chi^s_4+i\hat\chi^s_3)=p\cdot a,\nn\\
 &&G^s_2\equiv 2 \hat\chi^s_2-i\hat\chi^s_5,\nn\\
 &&G^s_5 \equiv\hat\chi^s_6=p\cdot\psi.
 \e{727}
 The basic ansatz for the states are here (cf \rl{711})
 \be
   &&|\psi\hb=\sum_{n=-\infty}^{\infty}\sum_{k=1}^\infty\sum_{\al=1}^4{1\over k!}A^{(n)}_{\al\mu_1\mu_2\cdots\mu_k}(x)|\al,n\hb^{\mu_1\mu_2\cdots\mu_k}+\!\!\!\sum_{n=-\infty}^{\infty}\sum_{\al=1}^4\phi_{\al}^{(n)}(x)|\al,n\hb,\nn\\
  &&|\al,n\hb^{\mu_1\mu_2\cdots\mu_k}\equiv a^{\mu_1\dag}a^{\mu_2\dag}\cdots a^{\mu_k\dag}|\al,n\hb, \quad |\al,n\hb\equiv|\al\hb|0\hb|0\hb_p |n\hb, 
    \e{72701}
    where $\al$ is a spinor index. $|\al\hb$ is a spinor state built from $\psi^{\mu}$ (see \eg the appendix in \cite{Marnelius:1989br}).
    
    $G^s_r|\psi\hb=0$ lead to the same equations as in subsection 7.2 except that all fields here have a spinor index. In addition $G^s_5|\psi\hb=0$ implies the Dirac equations 
    \be
\ga^{\nu}\dif_{\nu}\phi^{(n)}(x)=0,\quad\ga^{\nu}\dif_{\nu}A^{(n)}_{\mu_1\dots\mu_k}(x)=0,\;\;k\geq1.
\e{72702}

 \subsection{Quantization of the simple free pseudoclassical\\ $\Xi=0$ model} 
The Gupta-Bleuler quantization for the simple free pseudoclassical model with $\Xi=0$ (considered in section 6 and appendix B.4) is based on the constraint operators
\be
 &&G^s_0 \equiv 2\hat\chi^s_1= p^2,\nn\\
 &&G^s_1 \equiv \frac{1}{\sqrt 2}(\hat\chi^s_4+i\hat\chi^s_3)=p\cdot a,\nn\\
 &&G^s_5 \equiv\hat\chi^s_6=p\cdot\psi,
\e{728}
with the algebra
\be
&&[G^s_5,G^s_5]=G^s_0.
\e{729}
The state ansatz is also here given by \rl{72701}. The resulting equations are then the ones in \rl{721} with a spinor index on the wave functions together with the Dirac like equations \rl{72702}.

 \subsection{Quantization of the extended free pseudoclassical $\Xi=0$ model} 
Next we turn to the free extended pseudoclassical model with $\Xi=0$ considered in section 4.1 (see also appendix B.5). The constraint operators for a Gupta-Bleuler quantization may here be chosen to be
 \be
 &&G^s_0 \equiv 2\hat\chi^s_1= p^2,\nn\\
 &&G^s_1 \equiv \frac{1}{\sqrt 2}(\hat\chi^s_4+i\hat\chi^s_3)=p\cdot a,\nn\\
 &&G^s_3 \equiv \hat\chi^s_2+\hat\chi^s_8=a^{\dag}a+2-{1\over2}(e^2+\omega^2),\nn\\
  &&G^s_4 \equiv\frac{1}{2}(\hat\chi^s_8-\hat\chi^s_2+i\hat\chi^s_5)=a^2+\frac{1}{2}(e^2-\omega^2)-\frac{i}{2}(\omega e+e\omega), \nn\\
&&G^s_5 \equiv\hat\chi^s_6=p\cdot\psi,\nn\\
&&G^s_6 \equiv{1\over\sqrt2}(\hat\chi^s_9+i\hat\chi^s_7)=a\cdot\psi+{1\over\sqrt2}(\omega\theta+ie\theta),
 \e{731}
satisfying the Lie algebra,
 \be
&& [G^s_1,G^s_3]=G^s_1, \quad  [G^s_3,G^s_4]=-2G^s_4, \quad  [G^s_5,G^s_5]=2G^s_0,\nn\\
 && [G^s_5,G^s_6]=G^s_1,\quad  [G^s_6,G^s_6]=2G^s_4, \quad  [G^s_6,G^s_3]=G^s_6.
 \e{732}
In order to quantize this pseudoclassical model we need   a  representation of the odd hermitian operator $\theta$ in $G^s_6$ introduced in section 4.1.
We may write such a state representation in terms of one Grassmann even $|\;\hb_e$ and one Grassmann odd state $|\;\hb_o$ related by either of the following two choices
\be
i)&& \qquad\theta|\;\hb_e=i|\;\hb_o, \quad \;\;\;\theta|\;\hb_o=i|\;\hb_e,\nn\\
ii)&&\qquad\theta|\;\hb_e=-i|\;\hb_o, \quad\theta|\;\hb_o=-i|\;\hb_e .
\e{7271}
Let us focus on the first choice and let 
\be
&&(|\;\hb_e)^{\dag}={}_e\vb\;|, \quad (|\;\hb_o)^{\dag}={}_o\vb\;|, 
\e{7272}
and
\be
&&{}_e\vb\;|\;\hb_o={}_o\vb\;|\;\hb_e=0,\nn\\
&&{}_e\vb\;|\;\hb_e={}_o\vb\;|\;\hb_o={1\over2}.
\e{7273}
In terms of these states we define the two states $|i\hb$, $i=1,2$, by
\be
&&|1\hb\equiv|\;\hb_e+i|\;\hb_o,\nn\\
&&|2\hb\equiv|\;\hb_e-i|\;\hb_o,
\e{7274}
which from \rl{7273} satisfy the normalization
\be
&&\vb1|1\hb=\vb2|2\hb=1,\nn\\
&&\vb1|2\hb=\vb2|1\hb=0.
\e{7275}
We have then the following  matrix representation of the operator $\theta$,
\be
\vb i|\theta| j\hb\sim \left(\begin{array}{cc} 0 & 1 \\ -1 & 0\end{array}\right).
\e{7276}
The state ansatz for the operators \rl{731} is here given by
 \be
   &&|\psi\hb=\sum_{n=-\infty}^{\infty}\sum_{k=1}^\infty\sum_{\al=1}^4\sum_{i=1}^2{1\over k!}A^{(n,i)}_{\al\mu_1\mu_2\cdots\mu_k}(x)|\al,n,i\hb^{\mu_1\mu_2\cdots\mu_k}+\nn\\&&\quad \quad \quad+\!\!\!\sum_{n=-\infty}^{\infty}\sum_{\al=1}^4\sum_{i=1}^2\phi_{\al}^{(n,i)}(x)|\al,n,i\hb,\nn\\
  &&|\al,n,i\hb^{\mu_1\mu_2\cdots\mu_k}\equiv a^{\mu_1\dag}a^{\mu_2\dag}\cdots a^{\mu_k\dag}|\al,n,i\hb, \nn\\ &&|\al,n,i\hb\equiv|\al\hb|0\hb|0\hb_p |n\hb|i\hb.
    \e{7277}
$G_5|\psi\hb=0$  yields as before the massless Dirac like equation 
\be
&&\ga^{\nu}\dif_{\nu}\phi^{(n,i)}=0,\quad\ga^{\nu}\dif_{\nu}A_{\mu_1\dots\mu_k}^{(n,i)}=0,\quad k\geq1,
\e{733}
and in addition to the equations found before in \rl{721}, \rl{725} and \rl{726} with spinor and $i$-indices we also have from $G_6|\psi\hb=0$ the equations (suppressing spinor indices):
\be
&&\ga^{\nu} A^{(n,1)}_{\nu}-(n+1)\phi^{(n+1,2)}+\phi^{(n-1,2)}=0,\nn\\
&&\ga^{\nu} A^{(n,2)}_{\nu}+(n+1)\phi^{(n+1,1)}-\phi^{(n-1,1)}=0, \nn\\
&&\ga^{\nu} A^{(n,1)}_{\nu\mu_1\cdots\mu_k}-(n+1)A^{(n+1,2)}_{\mu_1\cdots\mu_k}+A^{(n-1,2)}_{\mu_1\cdots\mu_k}=0,\nn\\
&&\ga^{\nu} A^{(n,2)}_{\nu\mu_1\cdots\mu_k}+(n+1)A^{(n+1,1)}_{\mu_1\cdots\mu_k}-A^{(n-1,1)}_{\mu_1\cdots\mu_k}=0, \quad k\geq1.
\e{734}

\setcounter{equation}{0}
 \section{Conclusions}
 We have reviewed the classical derivations of Poincar\'e invariant massless representations first given by Wigner and Bargmann with particular emphasis on the continuous spin representation which we prefer to call the infinite spin representation or Wigner's $\Xi$-representation. We have then derived classical particle models from these representations in the spirit of the general procedure given in \cite{Marnelius:1990de}. For Wigner's $\Xi$-representation for integer spins we have \eg found a reparametrization invariant higher order geometrical theory whose Lagrangian with gauge fixed time essentially is the model once proposed by Zoller  \cite{Zoller:1994cl}.
 
 The mechanics of
 the derived models are rather peculiar since the velocity $\dox$ in general
 is space-like (which is manifest in the extended $\Xi=0$ model in section
 6). The models describe therefore tachyons. However, the models describe not
 normal tachyons since they have  light-like momenta $p$ due to the
 fact that  $p$ partly is proportional to the acceleration of the
 particle which also is peculiar.  Although these features do not prohibit the models from being consistent as free particle models, they do cause problems when we consider interactions. In fact, we have not found any consistent interactions with an external vector field and not with general gravity. However, consistent models may at least for $\Xi=0$ be defined on (anti)de Sitter space.  The interaction
 problems found here might be connected to the problems to construct
 interacting higher spin fields (\cite{Aragone:1979co, *Fradkin:1987on,*Bengtsson:1983cu,*Bengtsson:1983cu2, *Fang:1978ma,*Fang:1980ma}, see also
 \cite{Vasiliev:2004hi,*Vasiliev:2001pr,*Sorokin:2004in}
 for recent reviews). This remains to be investigated.

We propose that the free particle models may be consistently quantized.  The appropriate framework for this is the BRST quantization.  We believe that such a quantization is inconsistent with the Dirac quantization used in the present as well as the original derivations of the representations. We give two covariant treatments of the equations from the Dirac quantization which are  found inconsistent. (Negative results are also found in \cite{Abbott:1976ma,*Hirata:1977qu} using a noncovariant treatment.) We consider, therefore,  a Gupta-Bleuler quantization which we expect to be closer to a correct BRST treatment. In this way we have, indeed, found consistent sets of covariant equations for most models which look like reducible higher spin equations. A peculiar feature is that we have a dynamical einbein variable in the models. Since it is unclear how they should be treated and interpreted we have not analysed the resulting equations in detail. It is suggested that the einbein variable might be treated as an extra dimension in the fields.\\ \\ \\

\noindent
{\bf Acknowledgement:} We would like to thank Lars Brink for initiating this work.

 \begin{appendix}
\newpage
\small
{
\section{Constraints and their algebras in the considered classical models}
In this appendix we list the constraints and their Poisson algebras
(the nonzero part) for the
classical models of  infinite spin particles considered in the text.   

\subsection{The free theory (sect. 3.1)} 

\noindent
\begin{tabular}{|l l|}
\hline
Constraints & Poisson algebra \\
\hline
$\chi_1\equiv\half p^2 $ & $\{\chi_4, \chi_2\}=\chi_3$\\
$\chi_2\equiv\half\bigl(\pi^2-e^2\bigr) $&$\{\chi_4, \chi_3\}=2\chi_1$\\
$\chi_3\equiv p\cdot\pi$ &$\{\chi_5, \chi_2\}=2\chi_2$\\
$\chi_4\equiv p\cdot\xi-{1\over e}\Xi $&$\{\chi_5, \chi_4\}=-\chi_4$\\
$\chi_5\equiv \pi\cdot\xi-\omega e$ &$\{\chi_5, \chi_3\}=\chi_3$\\
\hline
\end{tabular}

\noindent
This is a consistent higher order model where the Lagrangian $L$ is
given in \rl{311} and the Hamiltonian $H=\chi_4$.

\subsection{Interaction with an external vector field $A_{\mu}(x)$\\*
  (sect. 5.1)}

\begin{tabular}{|l l|}
\hline
Constraints & Poisson algebra \\
\hline
$\chi^\vt_1\equiv\half (\La^\vt)^2+\frac{1}{4}F_{\mu\nu}(x)(\xi^{\mu}\pi^{\nu}-\xi^{\nu}\pi^{\mu})$ &$\{\chi^\vt_4, \chi^\vt_2\}=\chi^\vt_3$ \\
$\chi^\vt_2\equiv\half\bigl(\pi^2-e^2\bigr) $&$\{\chi^\vt_4, \chi^\vt_3\}=2\chi^\vt_1$\\
$\chi^\vt_3\equiv \La^\vt\cdot\pi$ &$\{\chi^\vt_5, \chi^\vt_2\}=2\chi^\vt_2$\\
$\chi^\vt_4\equiv \La^\vt\cdot\xi-{1\over e}\Xi
$&$\{\chi^\vt_5, \chi^\vt_4\}=-\chi^\vt_4$\\
$\chi^\vt_5\equiv \pi\cdot\xi-\omega e$ & $\{\chi^\vt_5,
\chi^\vt_3\}=\chi^\vt_3$\\
\hline
\end{tabular}

\noindent
where $\La^\vt_{\mu}\equiv p_{\mu}-A_{\mu}(x)$ and $F_{\mu\nu}\equiv
\dif_\mu A_\nu-\dif_{\nu} A_{\mu}$. $L$ is given in \rl{501} and $H=\chi_4^\vt$.
This model is not consistent since  the expressions
$\{\chi^\vt_1,\chi^\vt_4\}$ in \rl{505}, and
$\{\chi^\vt_1,\chi^\vt_3\}$ in \rl{5051} do not vanish for non-trivial $A_{\mu}(x)$.

\subsection{Interaction with an external gravitational field
  $g_{\mu\nu}(x)$ (sect. 5.2)}

\begin{tabular}{|l l|}
\hline
Constraints & Poisson algebra \\
\hline
$\chi^\gt_1\equiv\frac{1}{2}(\La^\gt_\rho
g^{\rho\sigma}\La^\gt_\sigma-\pi_{\rho}\pi_{\sigma}g^{\sigma\ga}R^\rho_{\;\al\beta\ga}\xi^\al\xi^\beta)$
&$\{\chi^\gt_4, \chi^\gt_2\}=\chi^\gt_3$ \\
$\chi^\gt_2\equiv\half\bigl(\pi_{\mu}g^{\mu\nu}\pi_{\nu}-e^2\bigr)$ &$\{\chi^\gt_4, \chi^\gt_3\}=2\chi^\gt_1$ \\
$\chi^\gt_3\equiv \La^\gt_{\mu}g^{\mu\nu}\pi_{\nu}$ &$\{\chi^\gt_5, \chi^\gt_2\}=2\chi^\gt_2$ \\
$\chi^\gt_4\equiv \La^\gt_{\mu}\xi^{\mu}-{1 \over e}\Xi$ &$\{\chi^\gt_5, \chi^\gt_4\}=-\chi^\gt_4$ \\
$\chi^\gt_5\equiv \pi_{\mu}\xi^{\mu}-\omega e$ &$\{\chi^\gt_5,
\chi^\gt_3\}=\chi^\gt_3$\\
\hline
\end{tabular}

\noindent
where $\La^\gt_{\mu}\equiv
p_{\mu}-\pi_{\nu}\Ga^{\nu}_{\mu\la}\xi^{\la}$. 
This model is not consistent since  the expressions
$\{\chi^\gt_1,\chi^\gt_4\}$ in \rl{520}, and
$\{\chi^\gt_1,\chi^\gt_3\}$ in \rl{5201} do not vanish for non-trivial
$g_{\mu\nu}(x)$. $L$ is given in
\rl{510} and $H=\chi_4^\gt$.

\subsubsection{$\Xi=0$ representation in (anti)de Sitter spacetime (sect. 5.3)}
In (anti)de Sitter spacetime, where
$R_{\mu\al\beta\ga}=K\bigl(g_{\al\beta}g_{\mu\ga}-g_{\al\ga}g_{\beta\mu}\bigr)$, the model A.3
is consistent for $\Xi=0$ since Dirac's  consistency condition $\dot{\chi}_1^\DS=0$ then is satisfied (see \rl{532}). In this case the above nonvanishing brackets become linear in the constraints.

\subsection{Simple free $\Xi=0$ representation (sect. 6)}

\noindent
\begin{tabular}{|l l|}
\hline
Constraints & Poisson algebra \\
\hline
$\chi_1\equiv\half p^2 $&$\{\chi_4, \chi_3\}=2\chi_1$\\
$\chi_3\equiv p\cdot\pi$ &\\
$\chi_4\equiv p\cdot\xi$&\\
\hline
\end{tabular}

\noindent
This is a consistent minimal model which, however,  is not a higher
order model. $L$ is given in \rl{604} and $H$ in \rl{602}. If we add a constraint like $\chi_2\equiv\half\bigl(\pi^2-1\bigr)$ the resulting model is also consistent. In a way it is contained in the consistent free model A.1 for $\Xi=0$ which is a higher order model.

\subsection{Extended free $\Xi=0$ representation (sect. 6)}

\noindent
\begin{tabular}{|l l l|}
\hline
Constraints & Poisson algebra &\\
\hline
$\chi_1\equiv\half p^2 $ & $\{\chi_4, \chi_2\}=\chi_3$&  $\{\chi_8, \chi_3\}=\chi_4$\\
$\chi_2\equiv\half\bigl(\pi^2-e^2\bigr) $&$\{\chi_4, \chi_3\}=2\chi_1$& $\{\chi_8, \chi_5\}=2\chi_8$\\
$\chi_3\equiv p\cdot\pi$ &$\{\chi_5, \chi_2\}=2\chi_2$&\\
$\chi_4\equiv p\cdot\xi $&$\{\chi_5, \chi_4\}=-\chi_4$&\\
$\chi_5\equiv \pi\cdot\xi-\omega e$ &$\{\chi_5, \chi_3\}=\chi_3$&\\
$\chi_8\equiv \half\bigl(\xi^2-\omega^2\bigr)$& $\{\chi_8, \chi_2\}=\chi_5$&\\
\hline
\end{tabular}

\noindent
This is a consistent model which, however, not seems to be derivable
from a higher order model. $L$ is given in \rl{608} and $H$ in \rl{6072}

%
%

\section{Constraints and their algebras in the considered pseudoclassical models}
Here we list the constraints and their Poisson algebras
(the nonzero part) for the various pseudoclassical models of  infinite
spin particles considered in the text. 
\subsection{The free theory (sect. 4.1)}

\begin{tabular}{|l l l|}
\hline
Constraints & Poisson algebra &\\
\hline
$\chi^\st_1\equiv\half p^2 $ & $\{\chi^\st_4, \chi^\st_2\}=\chi^\st_3$
&$\{\chi^\st_7,\chi^\st_4\}=-\chi^\st_6$ \\
$\chi^\st_2\equiv\half\bigl(\pi^2-e^2\bigr) $&$\{\chi^\st_4,
\chi^\st_3\}=2\chi^\st_1$&$ \{\chi^\st_7, \chi^\st_5\}=-\chi^\st_7 $\\
$\chi^\st_3\equiv p\cdot\pi$ &$\{\chi^\st_5, \chi^\st_2\}=2\chi^\st_2$ &$
\{\chi^\st_7,\chi^\st_7\}=-2i\chi^\st_2$ \\
$\chi^\st_4\equiv p\cdot\xi-{1\over e}\Xi $&$\{\chi^\st_5, \chi^\st_3\}=\chi^\st_3$ &\\
$\chi^\st_5\equiv \pi\cdot\xi-\omega e$ &$\{\chi^\st_5, \chi^\st_4\}=-\chi^\st_4$&\\
$\chi^\st_6\equiv p\cdot\psi$ & $\{\chi^\st_6, \chi^\st_6\}=-2i\chi^\st_1$&\\
$\chi^\st_7\equiv\pi\cdot\psi+e\theta$
&$\{\chi^\st_6,\chi^\st_7\}=-i\chi^\st_3$&\\
 \hline
\end{tabular}

\noindent
This is a consistent higher order model where the Lagrangian $L$ is given
in \rl{417} and $H=\chi_4^\st$.

\subsection{Interaction with an external vector field $A_{\mu}(x)$ \\*(sect. 5.1.1)}

\begin{tabular}{|l l l|}
\hline
Constraints & Poisson algebra &\\
\hline
$\chi^\sa_1\equiv\half ({\La^\sa})^2+\frac{1}{4}F_{\mu\nu}(x)(\xi^{\mu}\pi^{\nu}-\xi^{\nu}\pi^{\mu})$ &$\{\chi^\sa_4, \chi^\sa_2\}=\chi^\sa_3$
&$\{\chi^\sa_7,\chi^\sa_4\}=-\chi^\sa_6$ \\
$\chi^\sa_2\equiv\half\bigl(\pi^2-e^2\bigr) $&$\{\chi^\sa_4,
\chi^\sa_3\}=2\chi^\sa_1$&$ \{\chi^\sa_7, \chi^\sa_5\}=-\chi^\sa_7 $\\
$\chi^\sa_3\equiv {\La^\sa}\cdot\pi$ &$\{\chi^\sa_5, \chi^\sa_2\}=2\chi^\sa_2$ &$
\{\chi^\sa_7,\chi^\sa_7\}=-2i\chi^\sa_2$\\
$\chi^\sa_4\equiv {\La^\sa}\cdot\xi-{1\over e}\Xi $&$\{\chi^\sa_5, \chi^\sa_3\}=\chi^\sa_3$ &\\
$\chi^\sa_5\equiv \pi\cdot\xi-\omega e$ & $\{\chi^\sa_5, \chi^\sa_4\}=-\chi^\sa_4$& \\
$\chi^\sa_6\equiv {\La^\sa}\cdot\psi$ &$\{\chi^\sa_6, \chi^\sa_6\}=-2i\chi^\sa_1+R$&\\
$\chi^\sa_7\equiv \pi\cdot\psi
+e\theta$&$\{\chi^\sa_6,\chi^\sa_7\}=-i\chi^\sa_3$&\\
\hline
\end{tabular}

\noindent
where $\La^\sa_{\mu}\equiv p_{\mu}-A_{\mu}(x)$ and $F_{\mu\nu}\equiv \dif_\mu A_\nu-\dif_{\nu} A_{\mu}$.
This model is not consistent since the expressions $R$ above and $\{\chi^\sa_1,\chi^\sa_4\}$, 
$\{\chi^\sa_1,\chi^\sa_3\}$, $\{\chi^\sa_6,\chi^\sa_4\}$,
$\{\chi^\sa_1,\chi^\sa_6\}$, $\{\chi_1^\sa, \chi_7^\sa\}$, and $\{\chi_6^\sa, \chi_3^\sa\}$,  in
\rl{505},\rl{5051}, \rl{508}, and \rl{509} do not vanish for
non-trivial external field $A_{\mu}(x)$. $L$ is given in \rl{5053} and $H=\chi_4^\sa$.

\subsection{Interaction with an external gravitational field
  $g_{\mu\nu}(x)$ (sect. 5.2.1)}

\begin{tabular}{|l l l|}
\hline
Constraints & Poisson algebra &\\
\hline
${\chi}^{\sgt}_1\equiv\frac{1}{2}({\La}^{\sgt}_\rho
g^{\rho\sigma}{\La}^{\sgt}_\sigma-\pi_{\rho}\pi_{\sigma}g^{\sigma\ga}R^\rho_{\;\al\beta\ga}\xi^\al\xi^\beta$&\!\!\!\!\!\!$+\frac{i}{2}\psi^a\psi^bR_{ab\mu\nu}\xi^\mu
g^{\nu{\la}}\pi_{{\la}})$ & \\
${\chi}^{\sgt}_2\equiv\half\bigl(\pi_{\mu}g^{\mu\nu}\pi_{\nu}-e^2\bigr)$ &$\{\chi^\sgt_4, \chi^\sgt_2\}=\chi^\sgt_3$
&$\{\chi^\sgt_7,\chi^\sgt_4\}=-\chi^\sgt_6$  \\
${\chi}^{\sgt}_3\equiv {\La}^{\sgt}_{\mu}g^{\mu\nu}\pi_{\nu}$ &$\{\chi^\sgt_4,
\chi^\sgt_3\}=2\chi^\sgt_1$&$ \{\chi^\sgt_7, \chi^\sgt_5\}=-\chi^\sgt_7 $ \\
${\chi}^{\sgt}_4\equiv {\La}^{\sgt}_{\mu}\xi^{\mu}-{1\over
  e}\Xi$ & $\{\chi^\sgt_5, \chi^\sgt_2\}=2\chi^\sgt_2$ &$
\{\chi^\sgt_7,\chi^\sgt_7\}=-2i\chi^\sgt_2$\\
${\chi}^{\sgt}_5\equiv \pi_{\mu}\xi^{\mu}-\omega e$&$\{\chi^\sgt_5,
\chi^\sgt_3\}=\chi^\sgt_3$ & $\{\chi^\sgt_7,\chi^\sgt_6\}=-i\chi^\sgt_3$\\
${\chi}^{\sgt}_6\equiv {\La}^{\sgt}_{\mu}V^{\mu}_{a}\psi^{a}$ & $\{\chi^\sgt_5, \chi^\sgt_4\}=-\chi^\sgt_4$ &\\
${\chi}^{\sgt}_7\equiv \pi_{\mu}V^{\mu}_{a}\psi^{a}+e\theta$
&$\{\chi^\sgt_6, \chi^\sgt_6\}=-2i\chi^\sgt_1$
&\!\!\!\!\!\!\!\!\!\!\!\!$+iR_3$\\
\hline
\end{tabular}

\noindent
where ${\La}^{\sgt}_{\mu}\equiv
p_{\mu}-\pi_{\nu}\Ga^{\nu}_{\mu{\la}}\xi^{{\la}}-\frac{i}{2}\psi^a{\om}_{ab\mu}\psi^b$.
This is not a consistent model since $R_3$ above and 
the  Poisson brackets $\{\chi^\sa_1,\chi^\sa_4\}$, 
$\{\chi^\sgt_1,\chi^\sgt_3\}$, $\{\chi^\sgt_6,\chi^\sgt_4\}$,
$\{\chi^\sgt_1,\chi^\sgt_6\}$, $\{\chi_1^\sgt, \chi_7^\sgt\}$, and
$\{\chi_6^\sgt, \chi_3^\sgt\}$  in \rl{525}-\rl{528} do not vanish for
nontrivial external fields. $L$ is given in \rl{521} and $H=\chi_4^\sgt$.

\subsubsection{$\Xi=0$ representation in (anti)de Sitter spacetime (sect. 5.3)}

In (anti)de Sitter spacetime Dirac's consistency conditions in B.3 lead to the condition $\Xi=0$ and two new constraints: $\chi^\sDS_8\equiv \half\bigl(\xi^{\mu}g_{\mu\nu}\xi^{\nu}-\omega^2\bigr)$ and $\chi^\sDS_9\equiv \xi_{\mu}V^{\mu}_a\psi^{a}+\omega\theta$. The above nonvanishing Poisson brackets, as well as the remaining Poisson brackets involving $\chi^\sDS_8$ and $\chi^\sDS_9$,
become  linear expressions in the constraints. The model is a consistent higher order model!  (The gauge group is however not a Lie group.)

\subsection{Simple free $\Xi=0$ representation (sect. 6)}
\noindent
\begin{tabular}{|l l|}
\hline
Constraints & Poisson algebra \\
\hline
$\chi_1^\st\equiv\half p^2 $&$\{\chi_4^\st, \chi_3^\st\}=2\chi_1^\st$\\
$\chi_3^\st\equiv p\cdot\pi$ &$\{\chi_6^\st, \chi_6^\st\}=-2i\chi_1^\st$\\
$\chi_4^\st\equiv p\cdot\xi$&\\
$\chi_6^\st\equiv p\cdot\psi$&\\
\hline
\end{tabular}

\noindent
This is a consistent minimal model which, however,  is not a higher order model. If we add a constraint like $\chi_2^\st\equiv\half\bigl(\pi^2-1\bigr)$ the resulting model is also consistent. If we furthermore add $\chi_7^\st=\pi\cdot\psi+\tet$ we again have consistency.  In a way we have then arrived at a model contained in the consistent free model B.1 for $\Xi=0$, which is a higher order model.

\subsection{Extended free $\Xi=0$ representation (sect. 6)}

\begin{tabular}{|l l l l|}
\hline
Constraints & Poisson algebra & &\\
\hline
$\chi^\st_1\equiv\half p^2 $ & $\{\chi^\st_4, \chi^\st_2\}=\chi^\st_3$
&$\{\chi^\st_7,\chi^\st_7\}=-2i\chi^\st_2$ &$\{\chi^\st_9,\chi^\st_9\}=-2i\chi^\st_8$\\
$\chi^\st_2\equiv\half\bigl(\pi^2-e^2\bigr) $&$\{\chi^\st_4,
\chi^\st_3\}=2\chi^\st_1$&$\{\chi^\st_8,\chi^\st_2\}=\chi^\st_5$& $\{\chi^\st_8, \chi^\st_7\}=\chi^\st_9$\\
$\chi^\st_3\equiv p\cdot\pi$ &$\{\chi^\st_5, \chi^\st_2\}=2\chi^\st_2$ & $\{\chi^\st_8,\chi^\st_3\}=\chi^\st_4$&\\
$\chi^\st_4\equiv p\cdot\xi$&$\{\chi^\st_5, \chi^\st_3\}=\chi^\st_3$ &$\{\chi^\st_8,\chi^\st_5\}=2\chi^\st_8$&\\
$\chi^\st_5\equiv \pi\cdot\xi-\omega e$ &$\{\chi^\st_5, \chi^\st_4\}=-\chi^\st_4$&$\{\chi^\st_9,\chi^\st_2\}=\chi^\st_7$&\\
$\chi^\st_6\equiv p\cdot\psi$ & $\{\chi^\st_6, \chi^\st_6\}=-2i\chi^\st_1$&$\{\chi^\st_9,\chi^\st_3\}=\chi^\st_6$&\\
$\chi^\st_7\equiv\pi\cdot\psi+e\theta$
&$\{\chi^\st_6,\chi^\st_7\}=-i\chi^\st_3$&$\{\chi^\st_9,\chi^\st_5\}=\chi^\st_9$&\\
$\chi^\st_8\equiv \half\bigl(\xi^2-\omega^2\bigr)$&$\{\chi^\st_7,\chi^\st_4\}=-\chi^\st_6$ &$\{\chi^\st_9,\chi^\st_6\}=-i\chi^\st_4$&\\
$\chi^\st_9\equiv \xi_{\mu}\psi^{\mu}+\omega\theta$&$ \{\chi^\st_7, \chi^\st_5\}=-\chi^\st_7 $ &$\{\chi^\st_9,\chi^\st_7\}=-i\chi^\st_5$&\\
 \hline
\end{tabular}

\noindent
This is a consistent model which, however, not seems to be derivable from a higher order model.  (It looks like a flat, free version of B.3.1.)

}
\end{appendix}

\bibliographystyle{utphysmod2}
\bibliography{biblio1}

\providecommand{\href}[2]{#2}\begingroup\raggedright\begin{thebibliography}{10}

\bibitem{Wigner:1939on}
E.~P. Wigner,  {\em On unitary representations of the inhomogeneous {L}orentz
  group}, Annals Math. {\bf 40}, 149
(1939).

\bibitem{Bargmann:1948gr}
V.~Bargmann and E.~P. Wigner,  {\em Group theoretical discussion of
  relativistic wave equations}, Proc. Natl. Acad. Sci. ({USA}) {\bf 34}, 211
  (1948).

\bibitem{Wigner:1963in}
E.~P. Wigner,  {\em Invariant quantum mechanical equations of motion}, in
  Theoretical {P}hysics, International {A}tomic {E}nergy {A}gency, Vienna 59
(1963).

\bibitem{Ostrogradski:1850mv}
M.~V. Ostrogradski{,} Met. de l'Acad. de St.-Pet. {\bf 6}, 385 (1850).

\bibitem{Whittaker:1937an}
E.~T. Whittaker, {\em Analytical {D}ynamics}.
\newblock Cambridge Univ. Press, Cambridge, England, fourth~ed., 1937.

\bibitem{Lanczos:1970va}
C.~Lanczos, {\em The Variational Principles of Mechanics}.
\newblock Dover, fourth~ed., 1970.

\bibitem{Zoller:1994cl}
D.~Zoller,  {\em A classical theory of continuous spin and hidden gauge
  invariance}, Class. Quant. Grav. {\bf 11}, 1423
(1994).

\bibitem{Zoller:1990in}
D.~Zoller,  {\em Inconsistency of scale invariant curvature coupled to
  gravity}, Phys. Rev. Lett. {\bf 65}, 2236
(1990).

\bibitem{Savvidy:2003co}
G.~K. Savvidy,  {\em Conformal invariant tensionless strings}, Phys. Lett. {\bf
  B552}, 72
(2003).

\bibitem{Savvidy:2003te}
G.~K. Savvidy,  {\em Tensionless strings: Physical fock space and higher spin
  fields},
\href{http://www.arXiv.org/abs/hep-th/0310085}{{\tt hep-th/0310085}}.

\bibitem{Antoniadis:2004ph}
I.~Antoniadis and G.~Savvidy,  {\em Physical fock space of tensionless
  strings},
\href{http://www.arXiv.org/abs/hep-th/0402077}{{\tt hep-th/0402077}}.

\bibitem{Nichols:2002ne}
A.~Nichols, R.~Manvelyan and G.~K. Savvidy,  {\em New strings with world-sheet
  supersymmetry}, Mod. Phys. Lett. {\bf A19}, 363 (2004)
[\href{http://www.arXiv.org/abs/hep-th/0212324}{{\tt hep-th/0212324}}].

\bibitem{Fierz:1939su}
M.~Fierz{,} Helv. Phys. Acta {\bf 12}, 3 (1939).

\bibitem{Fierz:1939on}
M.~Fierz and W.~Pauli,  {\em On relativistic wave equations for particles of
  arbitrary spin in an electromagnetic field}, Proc. Roy. Soc. Lond. {\bf
  A173}, 211
(1939).

\bibitem{Abbott:1976ma}
L.~F. Abbott,  {\em Massless particles with continuous spin indices}, Phys.
  Rev. {\bf D13}, 2291
(1976).

\bibitem{Hirata:1977qu}
K.~Hirata,  {\em Quantization of massless fields with continuous spin}, Prog.
  Theor. Phys. {\bf 58}, 652
(1977).

\bibitem{Marnelius:1993ge}
R.~Marnelius,  {\em General state spaces for {BRST} quantizations}, Nucl. Phys.
  {\bf B391}, 621
(1993).

\bibitem{Marnelius:1990de}
R.~Marnelius and U.~Mårtensson,  {\em Derivation of manifestly covariant
  quantum models for spinning relativistic particles}, Nucl. Phys. {\bf B335},
  395
(1990).

\bibitem{Marnelius:1989br}
R.~Marnelius and U.~Mårtensson,  {\em {BRST} quantization of free massless
  relativistic particles of arbitrary spin}, Nucl. Phys. {\bf B321}, 185
(1989).

\bibitem{Aragone:1979co}
C.~Aragone and S.~Deser,  {\em Consistency problems of hypergravity}, Phys.
  Lett. {\bf B86}, 161
(1979).

\bibitem{Fradkin:1987on}
E.~S. Fradkin and M.~A. Vasiliev,  {\em On the gravitational interaction of
  massless higher spin fields}, Phys. Lett. {\bf B189}, 89
(1987).

\bibitem{Bengtsson:1983cu}
A.~K.~H. Bengtsson, I.~Bengtsson and L.~Brink,  {\em Cubic interaction terms
  for arbitrary spin}, Nucl. Phys. {\bf B227}, 31
(1983).

\bibitem{Bengtsson:1983cu2}
A.~K.~H. Bengtsson, I.~Bengtsson and L.~Brink,  {\em Cubic interaction terms
  for arbitrarily extended supermultiplets}, Nucl. Phys. {\bf B227}, 41
(1983).

\bibitem{Fang:1978ma}
J.~Fang and C.~Fronsdal,  {\em Massless fields with half integral spin}, Phys.
  Rev. {\bf D18}, 3630
(1978).

\bibitem{Fang:1980ma}
J.~Fang and C.~Fronsdal,  {\em Massless, half integer spin fields in de
  {S}itter space}, Phys. Rev. {\bf D22}, 1361
(1980).

\bibitem{Vasiliev:2004hi}
M.~A. Vasiliev,  {\em Higher spin gauge theories in various dimensions},
  Fortsch. Phys. {\bf 52}, 702 (2004)
[\href{http://www.arXiv.org/abs/hep-th/0401177}{{\tt hep-th/0401177}}].

\bibitem{Vasiliev:2001pr}
M.~A. Vasiliev,  {\em Progress in higher spin gauge theories},
\href{http://www.arXiv.org/abs/hep-th/0104246}{{\tt hep-th/0104246}}.

\bibitem{Sorokin:2004in}
D.~Sorokin,  {\em Introduction to the classical theory of higher spins},
\href{http://www.arXiv.org/abs/hep-th/0405069}{{\tt hep-th/0405069}}.

\end{thebibliography}\endgroup
\end{document}